\documentclass{aa}  
\usepackage{graphicx}
\usepackage{txfonts}
\usepackage{hyperref}
\usepackage{mathtools}
\usepackage{xcolor}
\newcommand{\eq}[1]{\begin{equation}  #1 \end{equation}}

\newcommand{\eqa}[1]{\begin{align}   #1 \end{align}}

\newcommand{\br}[1]{\left( #1 \right)}
\newcommand{\bc}[1]{\left\{ #1 \right\}}
\newcommand{\bb}[1]{\left[ #1 \right]}

\newcommand{\nn}{\nonumber}

\newcommand{\dd}{{\rm d}}

\newcommand{\pr}{{\rm Pr}}

\makeatletter
\renewcommand*\aa@pageof{, page \thepage{} of \pageref*{LastPage}}
\makeatother
\begin{document} 

   \title{Self-calibration and robust propagation of photometric redshift distribution uncertainties in weak gravitational lensing}
   \titlerunning{Photometric redshift distribution uncertainties in weak lensing}


   \author{B. St{\"o}lzner\inst{1},
                                B. Joachimi\inst{1},
                        A. Korn\inst{1},
                        H. Hildebrandt\inst{2},
                        \and
                        A. H. Wright\inst{2}
                        }
   \authorrunning{St{\"o}lzner et. al.}
   \institute{Department of Physics and Astronomy, University College London, Gower Street, London WC1E 6BT, UK\\ 
                                \email{benjamin.stolzner.18@ucl.ac.uk}
                        \and
                        Ruhr University Bochum, Faculty of Physics and Astronomy, Astronomical Institute (AIRUB), German Centre for Cosmological Lensing, 44780 Bochum, Germany
                        }

   \date{Received ; accepted }

 \abstract{We present a method that accurately propagates residual uncertainties in photometric redshift distributions into the cosmological inference from weak lensing measurements. The redshift distributions of tomographic redshift bins are parameterised using a flexible modified Gaussian mixture model. We fit this model to pre-calibrated redshift distributions and implement an analytic marginalisation over the potentially several hundred redshift nuisance parameters in the weak lensing likelihood, which is demonstrated to accurately recover the cosmological posterior. By iteratively fitting cosmological and nuisance parameters arising from the redshift distribution model, we perform a self-calibration of the redshift distributions via the tomographic cosmic shear measurements. Our method is applied to KV450 data, which comprises a combination of the third data release of the Kilo-Degree Survey and the VISTA Kilo-Degree Infrared Galaxy Survey. We find constraints on cosmological parameters that are in very good agreement with the fiducial KV450 cosmic shear analysis and investigate the effects of the more flexible model on the self-calibrated redshift distributions. We observe posterior shifts in the medians of the five tomographic redshift distributions of up to $\Delta z \approx 0.02$, which are, however, degenerate with an observed decrease in the amplitude of intrinsic galaxy alignments of about $10\%$.}
 


   \maketitle
%

\section{Introduction}
Weak gravitational lensing by the large-scale structure of the Universe, known as cosmic shear, is a powerful probe of cosmology. Rapid progress is being made in this field thanks to current and upcoming dedicated surveys such as the Dark Energy Survey \cite[DES; ][]{DES1, DES2, DES4, DES5}, the Subaru Hyper Suprime-Cam \cite[HSC; ][]{HSC1, HSC2}, and the European Southern Observatory (ESO) Kilo-Degree Survey \cite[KiDS; ][]{kuijken19, asgari20}. These surveys allow us to test the predictions of the standard Lambda cold dark matter ($\Lambda$CDM) cosmological model by constraining the matter density and the amplitude of matter density fluctuations to unprecedented precision. 

The main observables of weak lensing experiments are distortions of the ellipticities of background galaxies. Due to the weak signal and the impact of noise on the ellipticity measurement, this effect is measured statistically from large samples of galaxies. In order to model the theoretical prediction of the observed signal, an accurate calibration of the source redshift distribution is required. Given the large number of sources in a typical weak lensing survey, a complete spectroscopic redshift measurement is infeasible, and therefore the redshift is estimated from photometry (see \citealt{2019NatAs...3..212S} for a review). 

Several methods of photometric redshift calibration have been developed, such as direct calibration with spectroscopic sub-samples that are, potentially after re-weighting, representative of the full sample \citep{Lima, Bonnett, hildebrandt17} and angular cross-correlation clustering measurements with spectroscopic reference samples that overlap in redshift \cite[e.g.][]{Newman08,MatthewsNewman, Menard, mcleod17}. These methods can be merged using hierarchical Bayesian models that combine photometry measurements of individual galaxies and clustering measurements with tracer populations in a robust way \citep{sanchez,alarcon}. Furthermore, the redshift distribution in weak lensing surveys can be self-calibrated to some extent from the data themselves \citep{zhang10, CFHTLens, schaan20}. However, it is not only crucial to adopt a calibration method that estimates the true redshift distribution as precisely as possible, but also to choose a model that is flexible enough to describe the redshift distribution accurately. Such a model then allows us to propagate uncertainties in the redshift distribution, which arise from the calibration, into the actual cosmic shear analysis. 
 
Examples of such flexible redshift distribution models are Gaussian mixture models \citep{hoyle19, leistedt19} and hierarchical logistic Gaussian processes \citep{2020MNRAS.491.4768R}, which are applied to calibrate redshift distributions of galaxy samples via cross-correlation clustering measurements with overlapping spectroscopic samples. Gaussian processes are non-parametric, that is, they are not limited by a functional form, and therefore they fulfil the condition of being able to accurately fit the redshift distribution. However, since the fit parameters of the Gaussian process are non-linear, implementing the Gaussian process in the weak lensing likelihood (with fit parameters acting as nuisance parameters) and subsequent marginalisation requires a carefully chosen kernel that needs to be adapted to the redshift distribution. As an alternative to  Gaussian processes, the redshift distribution can be parameterised using linear basis function models with a fixed number of parameters, so that we can readily apply an analytic marginalisation over nuisance parameters.
 
It is common to parameterise the uncertainty on the redshift distribution using a shift in the mean of the distribution \citep{hildebrandt18, hildebrandt20, HSC2, DES3, hoyle18}, which captures the effect of uncertainties in the redshift distribution on the weak lensing analysis to the first order \citep{amara}. However, with larger surveys and decreasing statistical uncertainties, the contribution of higher orders will become important \citep{wright_som}. Furthermore, this parameterisation has the disadvantage of introducing probability weights at negative redshift values. Therefore, it is particularly interesting to adopt redshift distribution models that capture arbitrary variations in the distribution.

In this paper we present such a flexible redshift distribution model with linear fit parameters, as well as a technique that provides an analytic marginalisation over nuisance parameters that originate from the redshift distribution calibration. We parameterise the redshift distribution of samples of galaxies as a `comb', that is, a modified Gaussian mixture model with fixed, equidistant separation between components, identical variance, and a fixed number of components. The amplitudes of each Gaussian component serve as fit parameters in the redshift distribution calibration. 

We implement this redshift distribution model in the weak lensing likelihood. Since the model is linear in the fit parameters, we can analytically marginalise over the fitted amplitudes. The advantage of this procedure is that we can use a large number of components to fit the redshift distribution, which gives the model enough flexibility to fit a potentially complex redshift distribution. At the same time, we do not increase the total number of free sampling parameters of the likelihood, so that it is still feasible to sample the likelihood without a significant increase in runtime. Additionally, the marginalisation method allows us to propagate correlations between all fit parameters of the redshift distribution into the likelihood. Thus, we incorporate the correlation between the redshift distributions of tomographic bins, which are induced by the calibration method, into the cosmic shear analysis. We then demonstrate our approach on the KiDS+VIKING (KV450) dataset comprising the ESO KiDS \citep{kuijken15, kuijken19,dejong15,dejong17} and the fully overlapping VISTA Kilo-Degree Infrared Galaxy Survey \citep[VIKING; ][]{2013Msngr.154...32E} on a survey area of 450 ${\rm deg}^2$.

In order to allow the cosmic shear measurement to self-calibrate the redshift distribution, we adopt a two-step calibration method. First, we fit the comb model to the redshift histograms of \cite{hildebrandt18}, which were calibrated with deep spectroscopic sub-samples. Second, we apply an iterative fitting method of both the cosmological and nuisance parameters originating from the redshift calibration. The best-fit nuisance parameters then represent a model of the redshift distribution that is calibrated with both deep spectroscopic catalogues and cosmic shear data. In contrast to the fiducial analysis of \cite{hildebrandt18}, this method takes the full variability in the redshift distributions into account. When sampling the weak lensing likelihood, we then marginalise analytically over the set of best-fit nuisance parameters.
 
The paper is structured as follows: In Sect. \ref{sec:comb} the redshift distribution model is described. The theoretical modelling of the cosmic shear signal with analytic marginalisation over nuisance parameters is presented in Sect. \ref{sec:likelihood}, and the cosmic shear self-calibration method of the redshift distributions is described in Sect. \ref{sec:calibration}. Results are presented in Sect. \ref{sec:results} and discussed in Sect. \ref{sec:discussion}.
 
\section{Redshift distribution model  }
\label{sec:comb}
We modelled the redshift distribution, $n^{(\alpha)}(z)$, of each tomographic bin, $\alpha$, as a comb, that is, a slightly modified Gaussian mixture with $N_z$ components per bin, with fixed, equidistant separation in redshift between the components, and with identical variance $\sigma_{\rm comb}^2$:
\eq{
\label{eq:nz}
n^{(\alpha)}(z) \coloneqq \sum_{i=1}^{N_z} A_i^\alpha\; {\cal K}\br{z; z_i, \sigma_{\rm comb}^2}\;,
}
where the only free parameters to be fitted are the amplitudes $A_i^\alpha$. The model is linear in the amplitudes, which allows us to apply an analytic marginalisation over nuisance parameters when sampling the weak lensing likelihood. We chose to model the normalised `teeth' of the comb as
\eq{
\label{eq:basis_function}
{\cal K} \br{z; z_i, \sigma_{\rm comb}^2} = \frac{z}{N(z_i, \sigma_{\rm comb})}\, \exp \bc{- \frac{(z-z_i)^2}{2 \sigma_{\rm comb}^2} }\;, 
}
with normalisation over the interval $\bb{0,\infty}$:
\eq{
N(z_i, \sigma) = \sqrt{\frac{\pi}{2}}\, z_i\, \sigma\; {\rm erfc} \br{-\frac{z_i}{\sqrt{2} \sigma}} + \sigma^2 \exp \bc{-\frac{z_i^2}{2 \sigma^2}}.
}
While this method does not depend on a particular choice of ${\cal K}$, this form has the advantage of ensuring $n^{(\alpha)}(0)=0$. The redshift distribution is normalised so that
\eq{
\label{eq:normalisation}
\sum_{i=1}^{N_z} A_i^\alpha = 1\;.
}
Using Eq. \eqref{eq:normalisation}, we write the amplitude of the $N_z$-th component in terms of the remaining $N_z -1$ amplitudes:
\eq{
\label{eq:z_component}
A_N^\alpha = 1- \sum_{i=1}^{N_z-1} A_i^\alpha.
}
Inserting Eq. \eqref{eq:z_component} back into Eq. \eqref{eq:nz}, we find
\eqa{
\label{eq:comb_nz}
n^{(\alpha)}(z) &= {\cal K}\br{z; z_{N_z}, \sigma_{\rm comb}^2} \\ \nn
& ~~~~+ \sum_{i=1}^{N_z-1} A_i^\alpha \bb{ {\cal K}\br{z; z_i, \sigma_{\rm comb}^2}  - {\cal K}\br{z; z_{N_z}, \sigma_{\rm comb}^2} } \\ 
&\coloneqq \sum_{i=1}^{N_z} A_i^\alpha n_i(z)\;,
}
where we redefined the amplitude $A_{N_z}^\alpha \equiv 1$ and
\eq{
n_i(z) ={\cal K}\br{z; z_i, \sigma_{\rm comb}^2}  - {\cal K}\br{z; z_{N_z}, \sigma_{\rm comb}^2}\cdot \left(1-\delta_{iN_z}\right)\,.
}
Since the amplitudes should be positive, it is convenient to define
\eq{
a_i^\alpha \coloneqq \ln A_i^\alpha\;
}
as the actual fit parameters. The final result of the redshift calibration procedure with data $\boldsymbol{d}_{\rm cal}$ is then
\eq{
\label{eq:posterior}
\pr \br{ \bc{a_i^\alpha} | \boldsymbol{d}_{\rm cal}} \approx {\cal N}\br{ a_i^\alpha; {a^*}_i^\alpha, \tens{\Sigma}_{\rm cal}}\;,
}
where the posterior is approximated by a multivariate Gaussian distribution with best-fit ${a^*}_i^\alpha$ and covariance $\tens{\Sigma}_{\rm cal}$.

The model developed in this section is a particular example of a linear basis function model, which is a class of models that involve linear combinations of fixed non-linear functions of the input variables \citep[see for example][]{bishop}. While the linear dependence on the model parameters simplifies the analysis of this class of models, it requires the choice of an appropriate number of basis function components. In this work we determine the number of components by repeatedly performing a fit to the observed data with a varying number of components and selecting the model that provides the best fit to the observed data. Alternatively, a common approach in regression problems is to turn to Bayesian frameworks, which provide methods of determining the model complexity. A Bayesian framework requires the specification of a prior on the model parameters, which can work similarly to penalty terms in regularised least-squares regression. In particular, so-called shrinkage priors (see \citealt{vanErp20} for a review) are used to reduce the size of coefficient estimates by shrinking them towards zero. Variables that correspond to coefficients that are shrunk exactly to zero drop out of the model. Therefore, assuming a shrinkage prior provides a method to reduce the dimensionality of a given model. Furthermore, the linear model can be related to Gaussian process models when imposing Gaussian priors on the basis function amplitudes \citep[see for example][]{bishop}.

\section{Theoretical modelling of the cosmic shear signal}
\label{sec:likelihood}
\subsection{Weak lensing model and likelihood}
It is standard practice to use two-point statistics of the gravitational shear as summary statistics in weak lensing studies. In this paper we employ the two-point shear correlation function between two tomographic bins. However, it is straightforward to apply the formalism to other two-point statistics, such as Complete Orthogonal Sets of E/B-Integrals \cite[COSEBIs; ][]{schneider10} and band power estimates derived from the correlation functions \citep{schneider02,becker16,vanUitert18}, since they are all linear functionals of the cosmic shear angular power spectrum. 

The two-point correlation function between two tomographic bins, $\alpha$ and $\beta$, is defined via
\eq{
\label{eq:xi}
\xi_\pm^{(\alpha\beta)}(\theta) = \frac{1}{2\pi}\int_0^{\infty}{\rm d}\ell\, \ell C_{\rm GG}^{(\alpha\beta)}(\ell)J_{0,4}(\ell\theta),
}
where $J_{0,4}(\ell\theta)$ are Bessel functions of the first kind and $C_{\rm GG}^{(\alpha\beta)}(\ell)$ is the angular weak lensing convergence power spectrum.
Using the Limber approximation, the angular power spectrum reads \citep{kaiser92}
\eq{
C_{\rm GG}^{(\alpha \beta)}(\ell) = \int_0^{\chi_{\rm H}} \dd \chi \frac{q^{(\alpha)}(\chi) q^{(\beta)}(\chi)}{f_K^2(\chi)} P_\delta\br{\frac{\ell+1/2}{f_K(\chi)},\chi},
}
where $P_\delta$ is the matter power spectrum and $f_K$, $\chi$, and $\chi_{\rm H}$ are the co-moving angular diameter distance, the co-moving radial distance, and the co-moving horizon distance, respectively. The lensing efficiency is given by
\eq{
q^{(\alpha)}(\chi) = \frac{3H_0^2\Omega_{\rm m}}{2c^2} \frac{f_K(\chi)}{a({\chi})} \int_\chi^{\chi_{\rm H}} \dd \chi^\prime n_\chi^{(\alpha)}(\chi') \frac{f_K(\chi^\prime-\chi)}{f_K(\chi^\prime)},
}
with $a(\chi)$ being the scale factor and $n^\alpha(z){\rm d}z = n_\chi^{(\alpha)}(\chi){\rm d}\chi$ being the distribution of galaxies in redshift bin $\alpha$. Since $q^{(\alpha)}$ is a linear functional of the corresponding redshift distribution, it is straightforward to extract the amplitudes of the redshift distribution model. We find
\eqa{
C_{\rm GG}^{(\alpha \beta)}(\ell) &= \sum_{i,j=1}^{N_z} A_i^\alpha A_j^\beta \int_0^{\chi_{\rm H}} \dd \chi \frac{q_i(\chi) q_j(\chi)}{f_K^2(\chi)} P_\delta\br{\frac{\ell+1/2}{f_K(\chi)},\chi}\;\\
&\coloneqq \sum_{i,j=1}^{N_z} A_i^\alpha A_j^\beta\; c^\prime_{ij}(\ell)\;,
}
where we defined $q_i(\chi)$ as the lensing efficiency of the i-th component of the redshift distribution model, as defined in Eq. \eqref{eq:comb_nz}. In the final equality we defined $c^\prime_{ij}(\ell)$, which is the angular weak lensing power spectrum for two Gaussian mixture components at $z_i$ and $z_j$, as redshift distributions. Using Eq. \eqref{eq:xi}, we can compute the two-point shear correlation function between two tomographic redshift bins via
\eqa{
\xi_{\rm GG}^{(\alpha\beta)}(\theta) &= \sum_{i,j=1}^{N_z} A_i^\alpha A_j^\beta \int_0^\infty \frac{\dd \ell\, \ell}{2 \pi}\,J _{0/4}(\ell \theta)\, c^\prime_{ij}(\ell)\\
&\coloneqq  \sum_{i,j=1}^{N_z} A_i^\alpha A_j^\beta x_\pm^{(ij)}(\theta)\, ,\label{eq:xi_comb}
}
where we defined the two-point correlation function of two Gaussian comb components, $x_\pm^{(ij)}(\theta)$.

The observed weak lensing signal does not correspond to $\xi_\pm^{(\alpha\beta)}$ directly, but is contaminated by correlations between intrinsic ellipticities of neighbouring galaxies, II, and correlations between intrinsic ellipticities of foreground galaxies and background galaxies, GI \citep{hirata04}:
\eq{
\xi_\pm = \xi_{\rm GG} + \xi_{\rm II}+ \xi_{\rm GI}.
}
We followed the method presented in \cite{hildebrandt17} to model the effects of intrinsic galaxy alignments using a `non-linear linear' model \citep{hirata04,bridle07,joachimi11}. The contributions of GI and II alignments to the two-point shear correlation function were calculated using Eq. \eqref{eq:xi} with the II and GI angular power spectra:
\eqa{
C_{\rm II}^{(\alpha\beta)} &=\int\dd\chi\,\frac{n^{(\alpha)}(\chi)n^{(\beta)}(\chi)}{f_K^2(\chi)}P_{\rm II}\left(\frac{\ell + 1/2}{f_K(\chi)},\chi\right),\\ 
C_{\rm GI}^{(\alpha\beta)} &=\int\dd\chi\,\frac{q^{(\alpha)}(\chi)n^{(\beta)}(\chi)+n^{(\alpha)}(\chi)q^{(\beta)}(\chi)}{f_K^2(\chi)}P_{\rm GI}\left(\frac{\ell + 1/2}{f_K(\chi)},\chi\right).
}
Again, we used the linear dependence on the redshift distribution to extract the amplitudes of the redshift distribution model in analogy to Eq. \eqref{eq:xi_comb}. 
The power spectra of intrinsic galaxy alignments, $P_{\rm II}$ and $P_{\rm GI}$, are related to the matter power spectrum $P_\delta$ via
\eqa{
&P_{\rm II}(k,z) = F^2(z) P_\delta(k,z)\label{eq:P_II}\\
&P_{\rm GI}(k,z) = F(z) P_\delta(k,z),\label{eq:P_GI}
}
with
\eq{
\label{eq:F_z}
F(z) = -A_{\rm IA}C_1\rho_{\rm crit} \frac{\Omega_{\rm m}}{D_+(z)}.
}
Here, $D_+(z)$ denotes the linear growth factor, $\rho_{\rm crit}$ is the critical density at redshift $z=0$, and $C_1$ is a fixed normalisation constant that is set such that $C_1\rho_{\rm crit} = 0.0134$ \citep{joachimi11}. The redshift-independent amplitude of intrinsic alignments, $A_{\rm IA}$, is left as the only free parameter, which is implemented as a sampled nuisance parameter in the weak lensing likelihood. 

In general, the Gaussian log-likelihood is defined as
\eq{
{\cal L}  = -\frac{1}{2} \chi^2 + {\rm const.} = \sum_{ij} \left(d_i - m_i\right) C^{-1}_{ij} \left(d_j - m_j\right) + {\rm const.}\;,
}
where $d_i$ and $m_i$ denote the observed data and the model prediction, respectively, with the inverse covariance $C^{-1}_{ij}$. Thus, the weak lensing log-likelihood reads
\eq{
\label{eq:likelihood}
{\cal L} = - \frac{1}{2} \sum_{l,\alpha,\beta, l', \alpha', \beta'} \Delta_l^{(\alpha \beta)}\; Z_{(l,\alpha,\beta) \,  (l',\alpha',\beta')}\; \Delta_{l'}^{(\alpha' \beta')} + {\rm const.}\;,
}
where the indices $\alpha$ and $\beta$ run over all unique combinations of tomographic redshift bins. The two-point correlation function is analysed in $\theta$-bins that are denoted by $l$. The inverse covariance, which is assumed to be cosmology independent, is given by $\tens{Z}$ with elements $Z_{(l,\alpha,\beta) \,  (l',\alpha',\beta')}$. We have defined
\eq{
\label{eq:Delta}
\Delta_l^{(\alpha \beta)} \equiv d_l^{(\alpha \beta)} - \sum_{i,j=1}^{N_z} A_i^\alpha A_j^\beta\; x_\pm^{(ij)}(\theta_l)\;,
}
where $d_l$ denotes the element of the observed data vector in $\theta$-bin $l$ at angular scale $\theta_l$ and the indices $i$ and $j$ count over all possible combinations of components of the redshift distribution model. We note that all cosmology dependence is in the $x_\pm^{(ij)}(\theta)$.

\subsection{Marginal likelihood}
\label{sec:marginalisation}
The goal is to analytically derive the likelihood of a weak lensing experiment, marginalised over the potentially large number of nuisance parameters originating from the redshift calibration. We denote the parameters over which we sample the posterior distribution by $\boldsymbol{p}_{\rm sam}$ and parameters that we analytically marginalise over by $\boldsymbol{p}_{\rm ana}$. In particular, $\boldsymbol{p}_{\rm sam}$ includes cosmological parameter as well as nuisance parameters that account for intrinsic alignments, baryon feedback, and additive shear bias. The parameters $\boldsymbol{p}_{\rm ana}$ are the collection of amplitude parameters $\bc{a_i^\alpha}$. We obtain
\eq{
\pr \br{\boldsymbol{d} | \boldsymbol{p}_{\rm sam}} = \int \dd^{N_{\rm ana}} p_{\rm ana}\; \pr \br{\boldsymbol{d} | \boldsymbol{p}_{\rm sam}, \boldsymbol{p}_{\rm ana}} \pr \br{\boldsymbol{p}_{\rm ana}}\;,
}
where the prior on analytically marginalised parameters is given, in this case, by the posterior of the fit to the redshift distribution defined in Eq. \eqref{eq:posterior}.

In the following we assume that the overall weak lensing likelihood is Gaussian. Moreover, we apply a Laplace approximation to the posterior in the sub-space spanned by the redshift nuisance parameters, that is, we effectively assume the posterior to be well represented by a Gaussian in this regime. As shown by \cite{taylor10}, we can always maximise the likelihood for non-Gaussian distributions so that the assumption of Gaussianity locally around the peak of the likelihood is justified. The marginalised log-likelihood,
\eq{
{\cal L}_{\rm marg}  \equiv -2 \ln \pr \br{\boldsymbol{d} | \boldsymbol{p}_{\rm sam}}\;,
}
is then given by \citet{bridle02} and \citet{taylor10}:
\eq{
\label{eq:marginal_likelihood}
{\cal L}_{\rm marg} = {\cal L}_{\rm fid} - \frac{1}{2} \boldsymbol{\cal L^\prime}^\tau  \bb{ {\cal L^{\prime\prime}} + 2 \tens{\Sigma}^{-1}_{\rm cal} }^{-1} \boldsymbol{\cal L^\prime} + \ln \det \br{ \mathbb{I} + \frac{1}{2} {\cal L^{\prime\prime}} \tens{\Sigma}_{\rm cal}}\;,
}
where $\mathbb{I}$ denotes the identity matrix and ${\cal L}_{\rm fid}$ is the log-likelihood evaluated at the best fit of the nuisance parameters,
\eq{
{\cal L}_{\rm fid} \equiv -2 \ln \pr \br{\boldsymbol{d} | \boldsymbol{p}_{\rm sam}, \boldsymbol{p}^*_{\rm ana}}\;.
}
The vector of derivatives of the log-likelihood with respect to the nuisance parameters $a_i^\alpha$ is denoted by $\boldsymbol{\cal L^\prime}$, and the Hessian matrix of second derivatives with respect to the nuisance parameters is denoted by ${\cal L^{\prime\prime}}$. Analytic expressions of these quantities are given in Appendix \ref{ap:derivatives}. All of these derivatives are to be evaluated at the best fit of the nuisance parameters $\boldsymbol{p}^*_{\rm ana}$. The covariance matrix of nuisance parameters, originating from the calibration of the photometric redshift distribution, is given by $\tens{\Sigma}_{\rm cal} $.
For the $N_{\rm bin}$ tomographic bins used in the analysis, $N_{\rm bin} \times N_z$ nuisance parameters are marginalised over (modulo those amplitudes fixed by the normalisation of the redshift distribution given in Eq. \eqref{eq:normalisation}).

To test the validity of the approximate marginalised likelihood, we could perform a short initial Markov chain Monte Carlo (MCMC) analysis of the full likelihood, as proposed by \cite{taylor10}. This method would allow us to identify potential non-Gaussianities. Any non-Gaussian parameters could then be removed from the analytic marginalisation and instead numerically marginalised over via MCMC. The downside of this method is that the initial MCMC run is computationally expensive, especially when the number of nuisance parameters is large. As an alternative to a full MCMC, we could instead sample the likelihood with a reduced set of nuisance parameters in order to validate the approximations made in the marginalised likelihood. By selecting different sets of nuisance parameters, for example those describing the tails of the redshift distribution, we could probe the likelihood in different regions of the parameter space. This would allow us to gradually test the assumption that the posterior distribution in the sub-space spanned by the nuisance parameters can be approximated by a Gaussian.
\section{Redshift distribution self-calibration}
\label{sec:calibration}
It is standard practice to include nuisance parameters $\delta z_i$ in the weak lensing likelihood \citep{DES3, HSC2, asgari20}, which linearly shift the whole redshift distribution of each tomographic bin with a prior that is derived from the calibration with external datasets. When sampling the likelihood, this then allows for a self-calibration of the redshift distribution with cosmic shear measurements through a shift in the mean of the redshift distributions within the allowed prior range. 

In this work we replaced the shifts in the mean with the amplitudes of the comb components. Thus, the model can accommodate more complex variations in the redshift distributions. This, however, comes at the cost of an increase in the number of nuisance parameters from 5 to $5 \times N_z$, where the number of Gaussian components per bin, $N_z$, is typically of order 30.  Given the dimensionality of the new nuisance parameter space, a sampling of nuisance parameters via MCMC methods becomes computationally prohibitive. Thus, we marginalised analytically over the uncertainties on the fitted amplitudes, as outlined in Sect. \ref{sec:marginalisation}. By doing so, we lose the ability, however, to self-calibrate the redshift distributions with cosmic shear data since the amplitudes no longer appear as free parameters in the likelihood. In order to retain the calibration of the redshift distribution with cosmic shear data, we performed an additional calibration step.

\begin{figure}
\centering
\includegraphics[width=\linewidth]{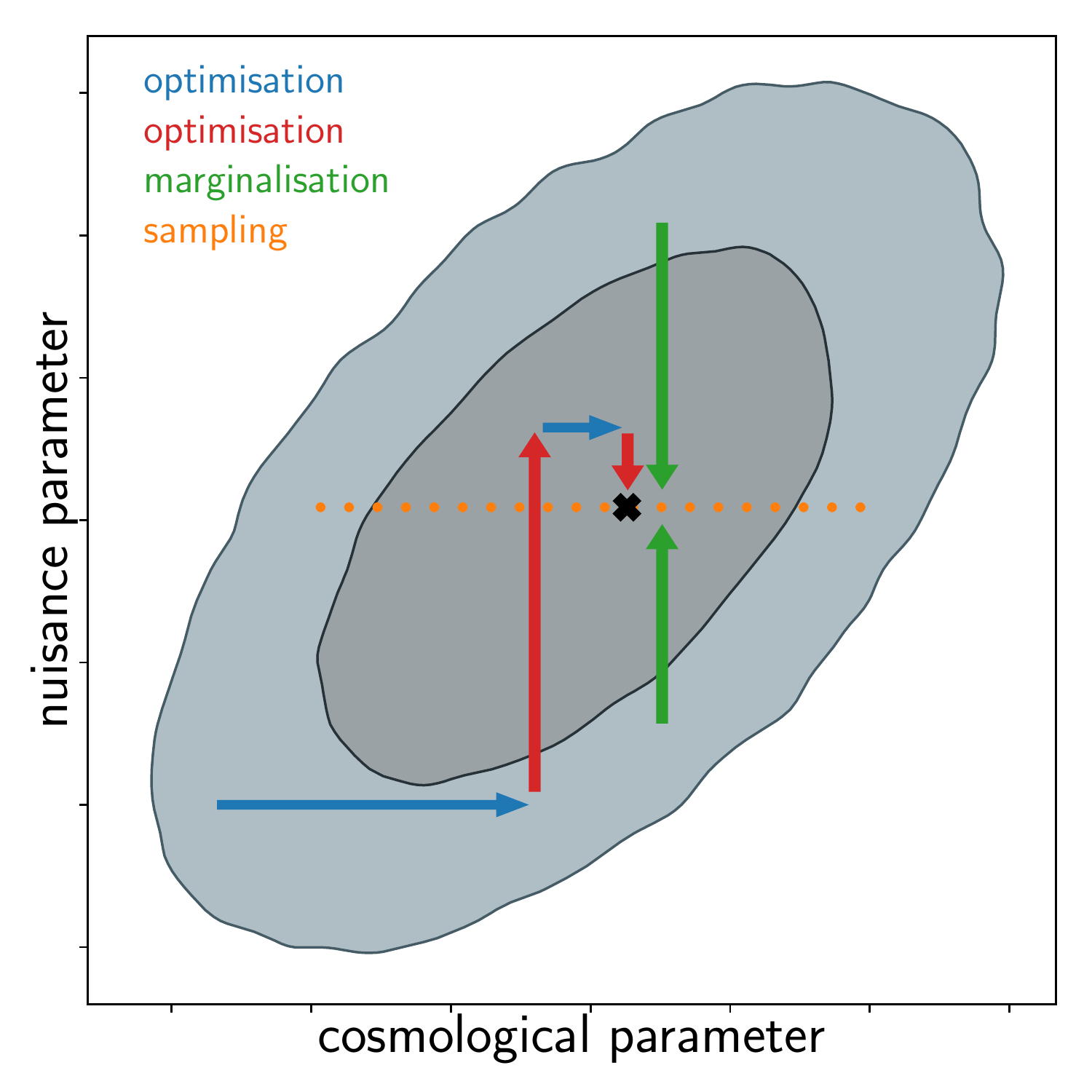}
\caption{Sketch of the iterative fitting method used to determine the best fit in the combined parameter space of cosmological and nuisance parameters. We alternate between optimising cosmological parameters (numerically; blue arrows), keeping nuisance parameters fixed, and optimising nuisance parameters (using Newton's method; red arrows), keeping cosmological parameters fixed. After several iterations we achieve convergence to the best fit in the combined parameter space. After optimising the likelihood, we set the amplitudes of the Gaussian comb to the best-fit parameters and proceed with sampling the likelihood in cosmological parameter space (dotted orange line) while analytically marginalising over nuisance parameters (green arrows).}
\label{fig:iterative_calibration}
\end{figure}

Our goal is to find the best fit in the combined parameter space of cosmological and nuisance parameters. Given the high dimensionality of the nuisance parameter space, we adopted an iterative method, which is illustrated in Fig. \ref{fig:iterative_calibration}:

First, we fitted the Gaussian comb model, defined in Sect. \ref{sec:comb}, to pre-calibrated redshift distribution histograms. This was done by minimising 
\eq{
\chi^2 = \sum_{ij} \left(n^{\rm data}_i - n^{\rm model}_i\right) C^{-1}_{ij} \left(n^{\rm data}_j - n^{\rm model}_j\right),
} 
where $n^{\rm data}_i$ and $n^{\rm model}_i$ are the observed and modelled histogram amplitudes in bin $i$, respectively, and $C^{-1}_{ij}$ denotes the inverse covariance matrix between the histogram bins in all five tomographic redshift bins. We estimated the uncertainties on the fit parameters by resampling the data vector using a multivariate Gaussian distribution, from which we calculated the covariance matrix $\Sigma_{\rm cal}$ of the fit parameters.

Second, we fixed the amplitudes of the Gaussian comb to the best-fit parameters found in the previous step. We then ran a non-linear optimiser to find the best-fit parameters $\boldsymbol{p}_{\rm sam}$ of the standard weak lensing likelihood conditioned on the best-fit parameters $\boldsymbol{p}_{\rm ana}$. This step is illustrated by the blue arrows in Fig. \ref{fig:iterative_calibration}.

Third, for fixed parameters $\boldsymbol{p}_{\rm sam}$, the displacement from the peak of the likelihood in the sub-space of parameters $\boldsymbol{p}_{\rm ana}$ is given by \citep{taylor10}:
\eq{
\label{eq:delta_p}
\delta\boldsymbol{p}_{\rm ana} = -\boldsymbol{\cal L^\prime}\left[{\cal L^{\prime\prime}}+2\tens{\Sigma}_{\rm cal}^{-1}\right]^{-1},
}
assuming a Gaussian prior on the parameters $\boldsymbol{p}_{\rm ana}$. Fixing the parameters $\boldsymbol{p}_{\rm sam}$ to the ones found in step 2, we used Newton's method to minimise Eq.\ \eqref{eq:delta_p} so that the parameters $\boldsymbol{p}_{\rm ana}$ converge towards the peak of the likelihood. Since the constraints on these parameters, which describe the redshift distribution, are dominated by the external priors through the original calibration, we anticipated the correction by the Newton step to be small. \footnote{For a likelihood that is close to Gaussian, we can find the maximum in one step. However, even if the initial redshift distribution is substantially different from the true underlying distribution, so that the likelihood at the initial values of the fit parameters is non-Gaussian, we can use Newton's method to iterate towards the peak \citep{taylor10}.} The red arrows in Fig. \ref{fig:iterative_calibration} represent this calibration step.  

By iterating over steps 2 and 3, we expected small corrections of both sets of parameters towards their best-fit values in the combined parameter space. The best-fit parameters $\boldsymbol{p}^*_{\rm ana}$ then represent the redshift distributions of each tomographic bin, calibrated using both spectroscopic catalogues and the actual cosmic shear measurements. After calibrating the redshift distributions, we set the amplitudes of the Gaussian comb in the weak lensing likelihood to the best-fit parameters and proceeded with the sampling of the likelihood in cosmological parameter space with pre-marginalised redshift distribution parameters. The sampling of the weak lensing likelihood is illustrated by the green arrows in Fig. \ref{fig:iterative_calibration}.

While it is advantageous to infer the initial values for $\boldsymbol{p}_{\rm ana}$ from a prior redshift calibration, the optimisation scheme itself is expected to be valid for any initial values. We tested for the existence of local minima in the posterior distribution by performing the optimisation for various choices of the initial values. We find that each optimisation converges towards consistent parameter values, which shows that the optimisation method succeeds in finding the global minimum of the posterior distribution.
\section{KV450 likelihood analysis}
\label{sec:results}
We used data from the ESO KiDS \citep{kuijken15, kuijken19,dejong15,dejong17} and the fully overlapping VIKING survey \citep{2013Msngr.154...32E}. This dataset, dubbed KV450, combines optical and near-infrared data on a survey area of 450 ${\rm deg}^2$. The photometric redshift calibration is greatly improved compared to the earlier KiDS dataset \citep{hildebrandt17} thanks to the addition of five near-infrared bands from VIKING that complement the four optical bands from KiDS. These additional bands improve the accuracy of photometric redshifts, which are used to define the tomographic bins. The fiducial technique of redshift calibration in KV450 utilised a weighted direct calibration, dubbed DIR, of five tomographic bins with deep spectroscopic catalogues. Uncertainties on the redshift distribution are estimated by a spatial bootstrapping method \citep{hildebrandt18}. The robustness of the photometric redshift calibration has been tested by excluding certain catalogues from the calibration sample as well as using alternative calibration samples. Additionally, the angular cross-correlation between KV450 galaxies and spectroscopic calibration samples has been studied as an alternative to the fiducial direct weighted calibration.

Our analysis is based on the fiducial KV450 cosmic shear analysis presented in \cite{hildebrandt18}, in which the combined KiDS+VIKING dataset \citep{2019A&A...632A..34W} is binned into five tomographic redshift bins based on their most probable Bayesian redshift, $z_{\rm B}$, inferred with the photo-z code {\sc BPZ} \citep{2000ApJ...536..571B}. Four bins of width $\Delta z = 0.2$ in the range $0.1 < z_{\rm B} \leq 0.9$ and a fifth bin with $0.9 < z_{\rm B} \leq 1.2$ are defined and calibrated using the aforementioned direct calibration method. The estimated redshift distribution is then used to model the two-point shear correlation function, and constraints on cosmological parameters are derived via sampling of the weak lensing likelihood. 

Self-organising maps (SOMs) have recently been proposed as a method to mitigate systematic biases arising from the redshift calibration, by assigning galaxies to groups based on their photometry \citep{buchs, wright_som, masters15}. This method allows samples of galaxies to be constrained such that they are fully represented by spectroscopic reference samples. It was recently applied to the KV450 \citep{wright_som_kv450} and KiDS-1000 \citep{hildebrandt20, asgari20} datasets. However, in those works the uncertainties on the redshift distributions are parameterised in terms of shifts in the mean of the redshift distributions with a prior that parameterises correlations between the redshift distributions of tomographic bins. This prior is inferred from simulations \citep{wright_som, hildebrandt20, vdBusch20}. A spatial bootstrapping was not performed, and as such an estimate of the full covariance of the redshift distribution is not available. In this work we therefore reverted to the fiducial KV450 dataset, for which such an estimation of the full covariance of the redshift distribution is available, and we leave the application to more recent KiDS datasets to future work.

In this work we calibrated the redshift distribution by fitting the Gaussian comb model defined in Sect. \ref{sec:comb} to the redshift distribution histograms of \cite{hildebrandt18}. Additionally, we extended the KV450 likelihood code originally used in the fiducial analysis of \cite{hildebrandt18} by implementing the analytic marginalisation over nuisance parameters. The original likelihood is publicly available in the {\sc MontePython}\footnote{\url{https://github.com/brinckmann/montepython_public}} package \citep{Audren:2012wb, Brinckmann:2018cvx}. We sampled the likelihood in the {\sc MultiNest} \footnote{\url{https://github.com/farhanferoz/MultiNest}} mode \citep{feroz09,feroz19} using the python wrapper {\sc PyMultiNest} \footnote{\url{https://github.com/JohannesBuchner/PyMultiNest}} \citep{buchner14}. The matter power spectrum is estimated with the public code {\sc Class}\footnote{\url{https://github.com/lesgourg/class_public}}\citep{blas11} with non-linear corrections from {\sc HMCode} \citep{mead15}. 

We adopted the cosmological model from \cite{hildebrandt18}, that is, a flat $\Lambda$CDM cosmology with five parameters: $\omega_{\rm CDM}$, $\omega_{\rm b}$, $A_{\rm s}$, $n_{\rm s}$, and $h$. Additionally, the model includes four nuisance parameters that account for intrinsic alignments ($A_{\rm IA}$), baryon feedback ($A_{\rm bary}$), and additive shear bias ($\delta c$ and $A_c$). We note that, in contrast to the fiducial KV450 analysis, we did not include linear shifts $\delta z_i$ in the mean redshift in each tomographic bin as nuisance parameters since variations in the redshift distributions are taken into account by the amplitudes of the Gaussian comb model with analytic marginalisation over the corresponding uncertainties. Our choices of priors for the nine cosmological and nuisance parameters are identical to the ones used in \cite{hildebrandt18} and are reported in Table \ref{tab:priors}. Finally, we adopted the cosmic shear data from \cite{hildebrandt18}, which consists of measurements of the two-point shear correlation functions between the five tomographic redshift bins and the corresponding analytic covariance matrix.
\begin{table}
\caption{Model parameters and their priors for the KV450 cosmic shear analysis, adopted from \cite{hildebrandt18}.}
\centering
\begin{tabular}{lll}
\hline\hline
Parameter & Symbol & Prior\\
\hline
CDM density & $\omega_{\rm CDM}$ & $[0.01, 0.99]$\\
Scalar spectrum amplitude & $\ln(10^{10}A_{\rm s})$ & $[1.7, 5.0]$\\
Baryon density & $\omega_{\rm b}$ & $[0.019, 0.026]$ \\
Scalar spectral index & $n_{\rm s}$ & $[0.7, 1.3]$ \\
Hubble parameter & $h$ & $[0.64, 0.82]$ \\
\hline
Intrinsic alignment amplitude & $A_{\rm IA}$ & $[-6, 6]$\\
Baryon feedback amplitude & $A_{\rm bary}$ & $[2.00, 3.13]$\\
Constant $c$-term offset & $\delta c$ & $0.0000\pm0.0002$ \\
2D $c$-term amplitude & $A_c$ & $1.01\pm0.13$\\
\hline
\end{tabular}
\tablefoot{The first five rows are cosmological parameters, and the remaining rows represent nuisance parameters. Brackets indicate top-hat priors, and values with errors indicate Gaussian priors. We note that, in contrast to \cite{hildebrandt18}, linear shifts in the mean of the redshift distributions are excluded.}
\label{tab:priors}
\end{table}
\subsection{Redshift distribution self-calibration}
\label{sec:redshift_calibration}
The first step in the calibration of the KV450 redshift distribution was to fit the modified Gaussian mixture model, defined in Sect. \ref{sec:comb}, to the redshift histograms of \cite{hildebrandt18}, which are pre-calibrated with deep spectroscopic samples. The fit was done simultaneously for all five tomographic redshift bins in order to account for the correlations between bins. We performed the fit using two different input data histograms: the fiducial histograms with a bin width of $\Delta z = 0.05$ and histograms with a smaller bin width of $\Delta z = 0.025$. Although both histograms trace the same underlying redshift distribution of galaxies in each tomographic bin, their biases and variances will generally be different. By fitting the comb model to the two types of histograms, we tested how the redshift distribution calibration is affected by noise. 

Moreover, we were free to choose the number of Gaussian components of the redshift distribution model and the variance of each component. In Appendix \ref{ap:calibration} we perform several tests, comparing different choices for the aforementioned free parameters, and address their impact on the cosmological analysis. In particular, as a rule of thumb for the width of the Gaussian component, $\sigma_{\rm comb}$, we limited ourselves to values that ensure an overlap of two to three components at each point in redshift space. We find that the analysis is robust with respect to these choices. In this section we report our fiducial result using a model with $N_z = 30$ equidistant components between $0\leq z \leq 2$ and a variance of $\sigma_{\rm comb} = 0.067$, which is equal to the separation between the mean redshift of each component. This model was fitted to the redshift histograms with a bin width of $\Delta z = 0.05$.
\begin{figure*}
\centering
\includegraphics[width=\linewidth]{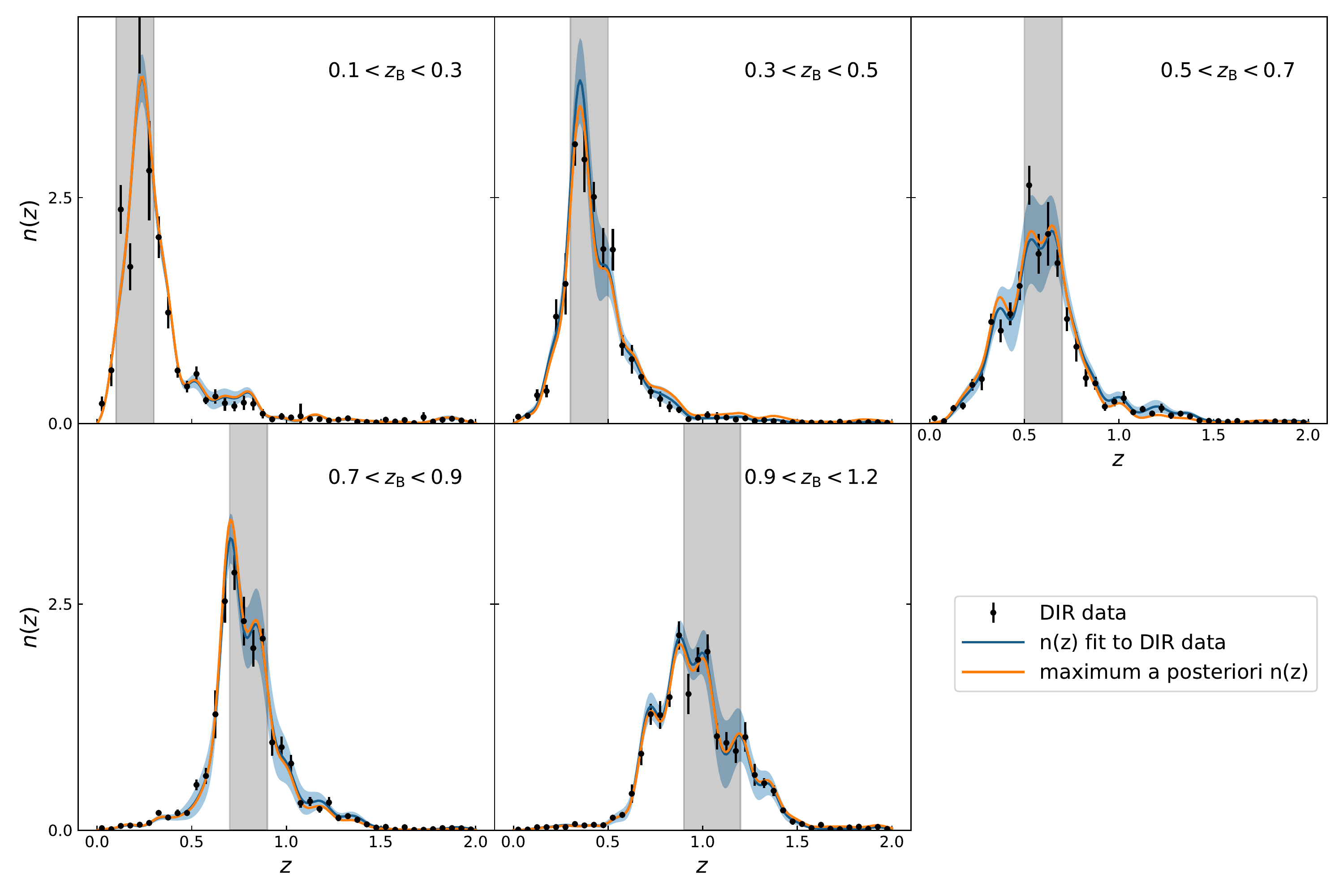}
\caption{Fit results of a Gaussian mixture with 30 components to the redshift distribution in five tomographic redshift bins. Blue curves indicate redshift distributions fitted to the pre-calibrated DIR redshift histograms, shown in black. Shaded regions indicate the uncertainties on the redshift distributions derived from the diagonal elements of the correlation matrix of fit parameters, shown in Fig. \ref{fig:correlation_matrix}. Orange curves represent the redshift distributions after iterative optimisation of cosmological and nuisance parameters.}
\label{fig:comb}
\end{figure*}
\begin{figure}
\centering
\includegraphics[width=\linewidth]{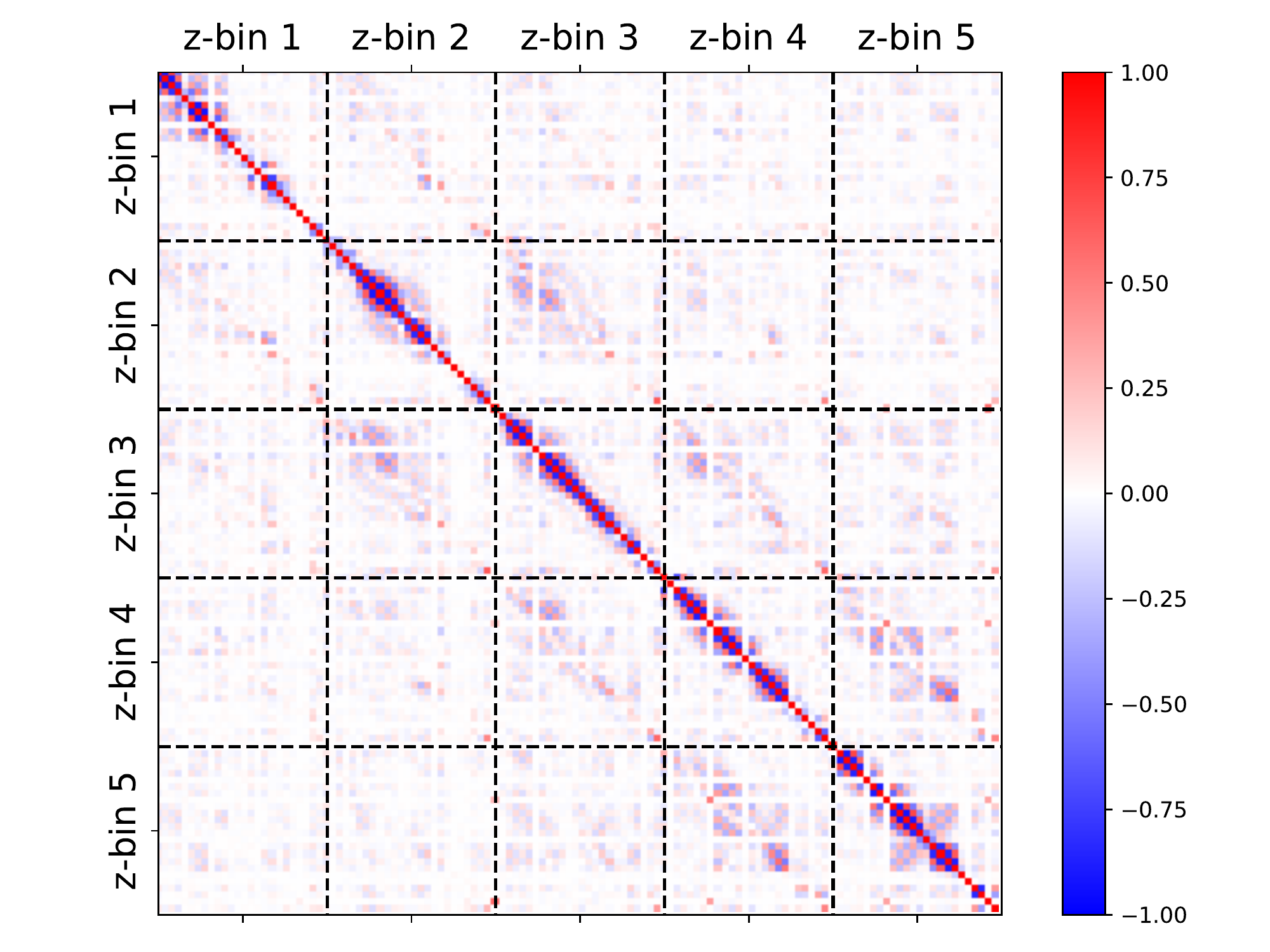}
\caption{Correlation matrix of best-fit comb amplitudes with 30 components per redshift bin.}
\label{fig:correlation_matrix}
\end{figure}
\begin{figure*}
\centering
\includegraphics[width=\linewidth]{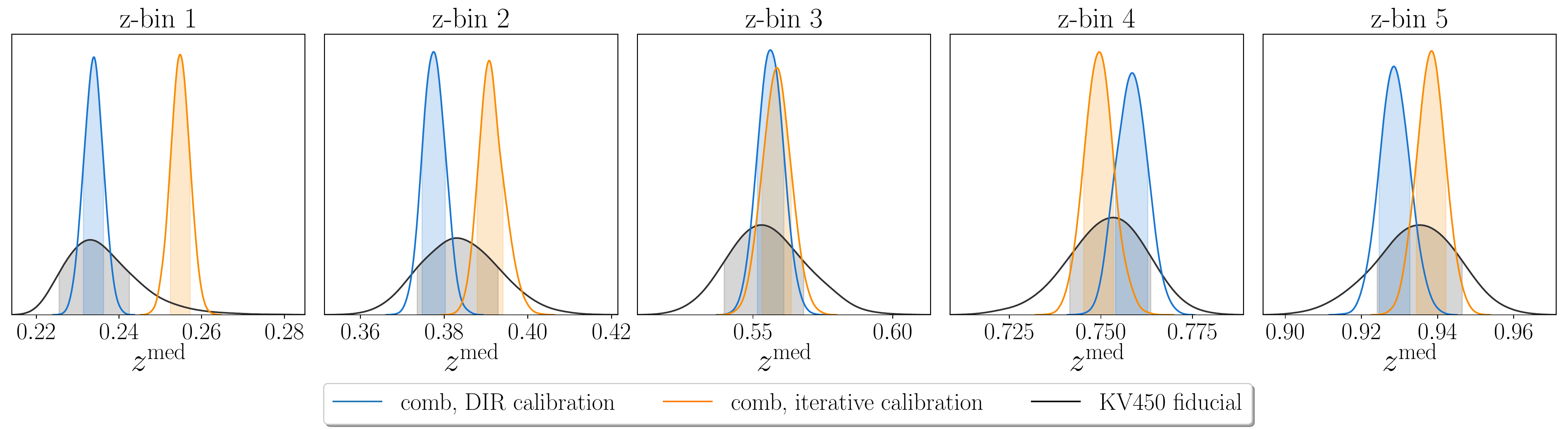}
\caption{Posterior distribution of the median redshift of each tomographic redshift bin, inferred by drawing realisations of the Gaussian comb amplitudes from a multivariate Gaussian distribution. Black curves indicate the median redshift of the KV450 redshift histograms calibrated using the fiducial DIR method. The blue curves show the median redshift of the Gaussian comb that is fitted to the DIR histograms. The orange curves represent the median redshift of the Gaussian comb after iterative self-calibration with cosmic shear measurements.}
\label{fig:medians_posterior}
\end{figure*}
The best-fit model is illustrated as blue curves in Fig. \ref{fig:comb}; the shaded regions indicate the uncertainties on the redshift distributions, which are derived from the diagonal elements of the covariance matrix of fit parameters. The correlation matrix of fit parameters is shown in Fig. \ref{fig:correlation_matrix}.  

We proceeded with a further calibration of the redshift distribution using the iterative fitting method of cosmological and nuisance parameters described in Sect. \ref{sec:calibration}. The fit result after each step for the two parameters that are mostly constrained by the data, the intrinsic alignment amplitude $A_{\rm IA}$, and the amplitude of matter density fluctuations, $S_8 = \sigma_8(\Omega_{\rm m}/0.3)^{0.5}$, are reported in Table \ref{tab:iterative_calibration}. We note that in this analysis, $S_8$ is a derived parameter that is inferred from {\sc Class}. 
\begin{table}
\caption{Results of the iterative fitting of cosmological and nuisance parameters to the KV450 cosmic shear data and comparison to the fiducial KV450 likelihood.}
\centering
\begin{tabular}{llll}
\hline\hline
& $A_{\rm IA}$ & $S_8$ & $\chi^2$\\
\hline
fiducial KV450 likelihood & $0.8656$  & $0.7708$ & $179.88$\\
\hline
1. cosmology optimisation & $0.7353$  & $0.7768$ & $180.67$\\
2. nuisance optimisation & ---  & --- & $179.11$\\
3. cosmology optimisation & $0.7903$  & $0.7882$ & $178.62$\\
\end{tabular}
\tablefoot{When optimising cosmological parameters, we fit $\omega_{\rm cdm}$, $\omega_{\rm b}$, $A_{\rm s}$, $n_{\rm s}$, and $h$ as well as the nuisance parameters $A_{\rm IA}$, $A_{\rm bary}$, $\delta c$, and $A_{\rm c}$. When optimising nuisance parameters, we vary the amplitudes of the Gaussian comb. Results are shown for the two most interesting parameters, $A_{\rm IA}$ and $S_8$, for which the cosmic shear likelihood has the largest constraining power. We find convergence after three iterations of the calibration, which results in a better fit to the cosmic shear data compared to the fiducial analysis.}
\label{tab:iterative_calibration}
\end{table}
The iterative optimisation method shows a fast convergence to the best fit in the full parameter space of cosmological and nuisance parameters after only two cosmology optimisation steps and one redshift nuisance parameter optimisation. This was unsurprising since we started from an already well-calibrated redshift distribution and as such expected only small corrections from the Newton optimisation step. 

Using the best-fit $\chi^2$ values as a measure of goodness of fit, we find that with $\chi^2 = 178.62$ our model provides an improvement in $\chi^2$ of roughly 1\% with respect to the fiducial KV450 model with a value of $\chi^2 = 179.88$. While in the present analysis allowing for a full variation in redshift distribution only gives a slight improvement compared to a linear shift in the mean, this method could become more relevant for future analyses with increased precision.

It is common practice to assess the goodness of fit by making the assumption that the $\chi^2$ statistic follows a $\chi^2$ distribution with $N_{\rm dof} = N_{\rm d} - N_\Theta$, where $N_{\rm d}$ is the size of the data vector and $N_\Theta$ is the number of sampling parameters. However, this assumption is only valid under the condition that the data are normally distributed, the model is linearly dependent on the sampling parameters, and there is no informative prior on the parameter ranges \citep[see for instance][]{joachimi20}. In general, these conditions are not met in cosmological analyses, and this is particularly true for this work since we assumed a Gaussian prior on the amplitudes of the Gaussian comb, which is inferred from the redshift distribution calibration. Therefore, the naive estimation of the number of degrees of freedom is a poor estimation of the true effective number of degrees of freedom since we added a large number of strongly correlated nuisance parameters with informative priors. For a conservative estimate of the number of degrees of freedom in our model, we can assume the nuisance parameters to be essentially fixed by the prior and therefore do not count them as sampling parameters, which leads to  $N_{\rm dof} = 186$. While a more robust estimate of the effective number of degrees of freedom can be inferred from mocks or posterior predictive data realisations \citep{spiegelhalter02, handley19, raveri19,joachimi20}, we refrain from a further interpretation of the goodness of fit.

We find a shift in the two most interesting parameters for which the cosmic shear likelihood has the largest constraining power, $A_{\rm IA}$ and $S_8$, compared to the fiducial KV450 analysis. These shifts are further investigated in the following section, where we sample the weak lensing likelihood and derive marginalised posteriors of cosmological parameters. The resulting redshift distributions after iterative self-calibration are illustrated as orange curves in Fig. \ref{fig:comb}. 

Figure \ref{fig:medians_posterior} shows comparisons of the median of the redshift distribution of each tomographic bin inferred from the original DIR histograms and the Gaussian comb before and after iterative calibration. We chose the median as our summary statistic since the mean of the DIR histograms is less stable with respect to variations in the cutoff redshift at the high-redshift tail of the distribution, which is most likely caused by the underestimation of the error bars in the DIR method. The median, on the other hand, is less sensitive to the choice of the cutoff redshift. We find that the fit of the Gaussian comb yields constraints on the median redshift that are about 50\% tighter   relative to the DIR histograms. Additionally, we observe that the shift in the median after iterative self-calibration is largest in the first two redshift bins and less significant in the three higher redshift bins. This is most likely caused by degeneracies between the amplitude of intrinsic alignments and the redshift distributions, which is discussed in the following sections. 

\subsection{Marginalisation over nuisance parameters} 
Using the redshift distribution calibrated in the previous section, we sampled the weak lensing likelihood in cosmological parameter space with analytical marginalisation over the uncertainty on the amplitudes of the fitted redshift distribution. 
Prior to the sampling of the marginalised likelihood, we tested whether we could reproduce the result of the fiducial KV450 analysis by sampling the likelihood with the comb model, but without applying the marginalisation over nuisance parameters. The results of this consistency test, discussed in Appendix \ref{ap:kv450_likelihood}, show that the two models are in good agreement. 
\begin{figure}
\centering
\includegraphics[width=\linewidth]{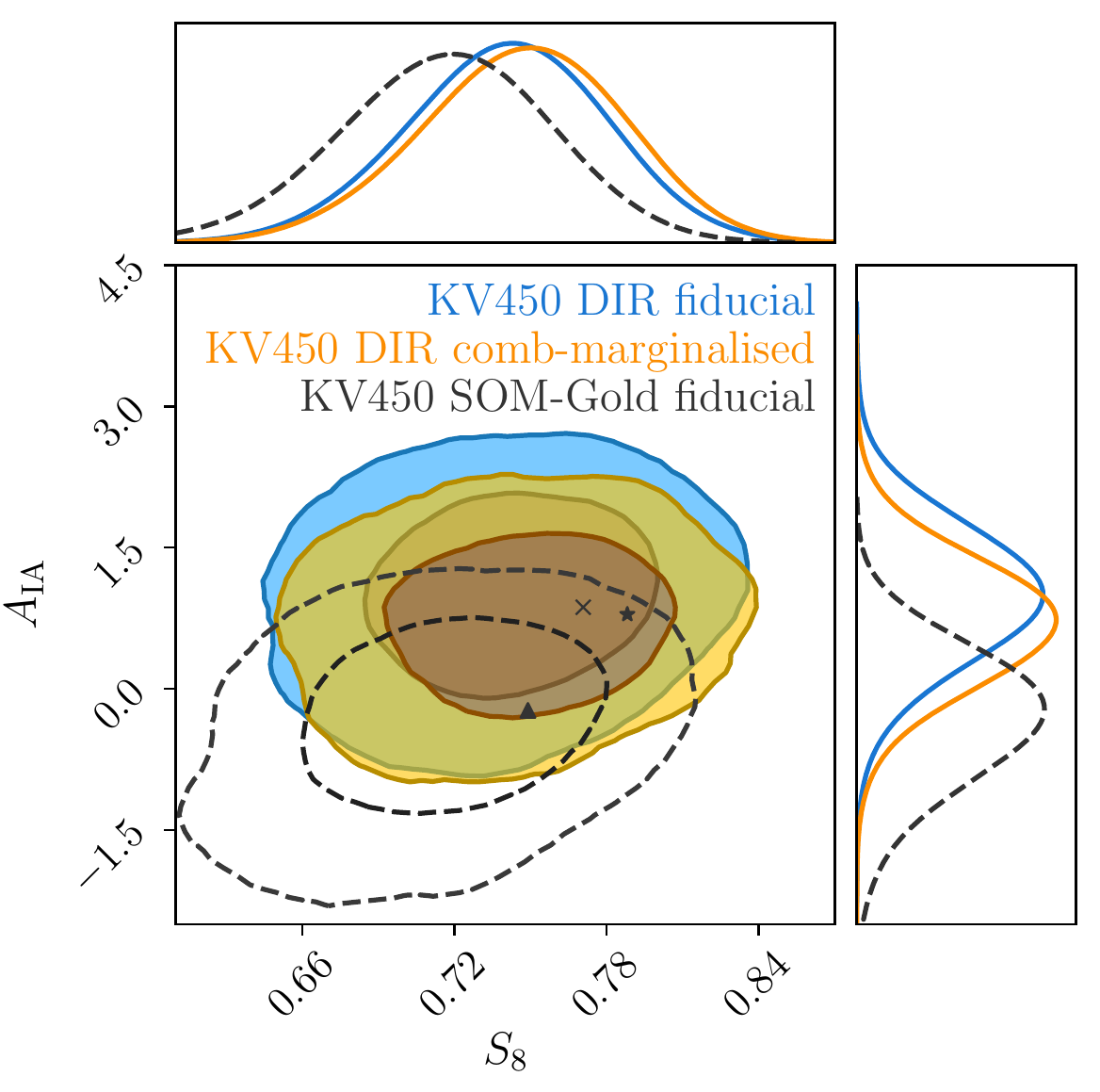}
\caption{Marginalised posteriors for $A_{\rm IA}$ and $S_8$. The orange contours present the results from the KV450 likelihood with a self-calibrated Gaussian comb and analytical marginalisation over nuisance parameters, while the blue contours refer to the fiducial KV450 constraints. The star indicates the best-fit values from Table \ref{tab:iterative_calibration} for the KV450 likelihood with a Gaussian comb, and the cross indicates the best-fit values for the fiducial KV450 likelihood. The dashed contour shows the posterior distribution from the KV450 `gold' sample \citep{wright_som_kv450}, which is constructed by removing photometric sources that are not directly represented by the overlapping spectroscopic reference samples using SOMs. Therefore, this contour is inferred from a different sample of galaxies with a different redshift distribution.}
\label{fig:AIA_vs_S8}
\end{figure}

Figure \ref{fig:AIA_vs_S8} illustrates the results, comparing (i) the KV450 likelihood with a Gaussian comb and analytical marginalisation over nuisance parameters with (ii) the fiducial KV450 likelihood. We show marginalised posteriors and best-fit values for the two parameters that are fully constrained with KV450 data, $A_{\rm IA}$ and $S_8$. The posterior distribution of all remaining parameters is shown in Appendix \ref{ap:posteriors}. We find a slight shift in the posterior towards smaller values of the intrinsic alignment amplitude and larger values of $S_8$. Additionally, Fig. \ref{fig:AIA_vs_S8} shows the posterior distribution for the KV450 `gold' sample, which is derived using SOMs \citep{wright_som_kv450}. We emphasise that \cite{wright_som_kv450} use a different selection of the photometric sample by removing photometric sources that are not directly represented by the overlapping spectroscopic reference samples. Thus, the redshift distributions of the fiducial KV450 sample and the KV450 gold sample are not comparable. 

The constraint from the KV450 gold sample on the intrinsic alignment amplitude, $A_{\rm IA}$, is compatible with $A_{\rm IA} = 0$, whereas \cite{hildebrandt18} found $A_{\rm IA} \approx 1$. However, these results are still consistent within their errors, as discussed by \cite{wright_som_kv450}. The iterative self-calibration of the redshift distribution performed in this work leads to a decrease in the intrinsic alignment amplitude of about $10\%$ (see Table \ref{tab:iterative_calibration}). Thus, we find a trend similar to that found by \cite{wright_som_kv450}, although the change in the intrinsic alignment amplitude is not as strong. Recent studies of intrinsic alignments have also found results that are in disagreement with the fiducial KV450 analysis, such as \cite{fortuna20}, who predict $A_{\rm IA} = 0.1 ^{+0.1}_{-0.1}$. Since the constraints on the intrinsic alignment amplitude differ between analyses and the role of intrinsic alignments is a subject of active research, it is worth investigating how this parameter can influence the theoretical prediction of the cosmic shear signal.

The intrinsic alignment amplitude, $A_{\rm IA}$, is not a cosmological parameter, but instead originates from the modelling of correlations between intrinsic ellipticities of neighbouring galaxies, II, and correlations between intrinsic ellipticities of foreground galaxies and background galaxies, GI. As can be inferred from Eqs. \eqref{eq:P_II} and \eqref{eq:P_GI}, the GI term gives a negative contribution to $\xi_\pm$ that is proportional to $A_{\rm IA}$, whereas the II term contributes positively, proportionally to $A_{\rm IA}^2$. Thus, a shift in the redshift distribution can (at least to some extent) be counteracted by a shift in the intrinsic alignment amplitude, so that overall we find a good fit to the observed cosmic shear two-point correlation function. This effect is a possible explanation for the observed shift in the contours in Fig. \ref{fig:AIA_vs_S8}, which nevertheless are still in good agreement. 

\begin{figure*}
\centering
\includegraphics[width=\linewidth]{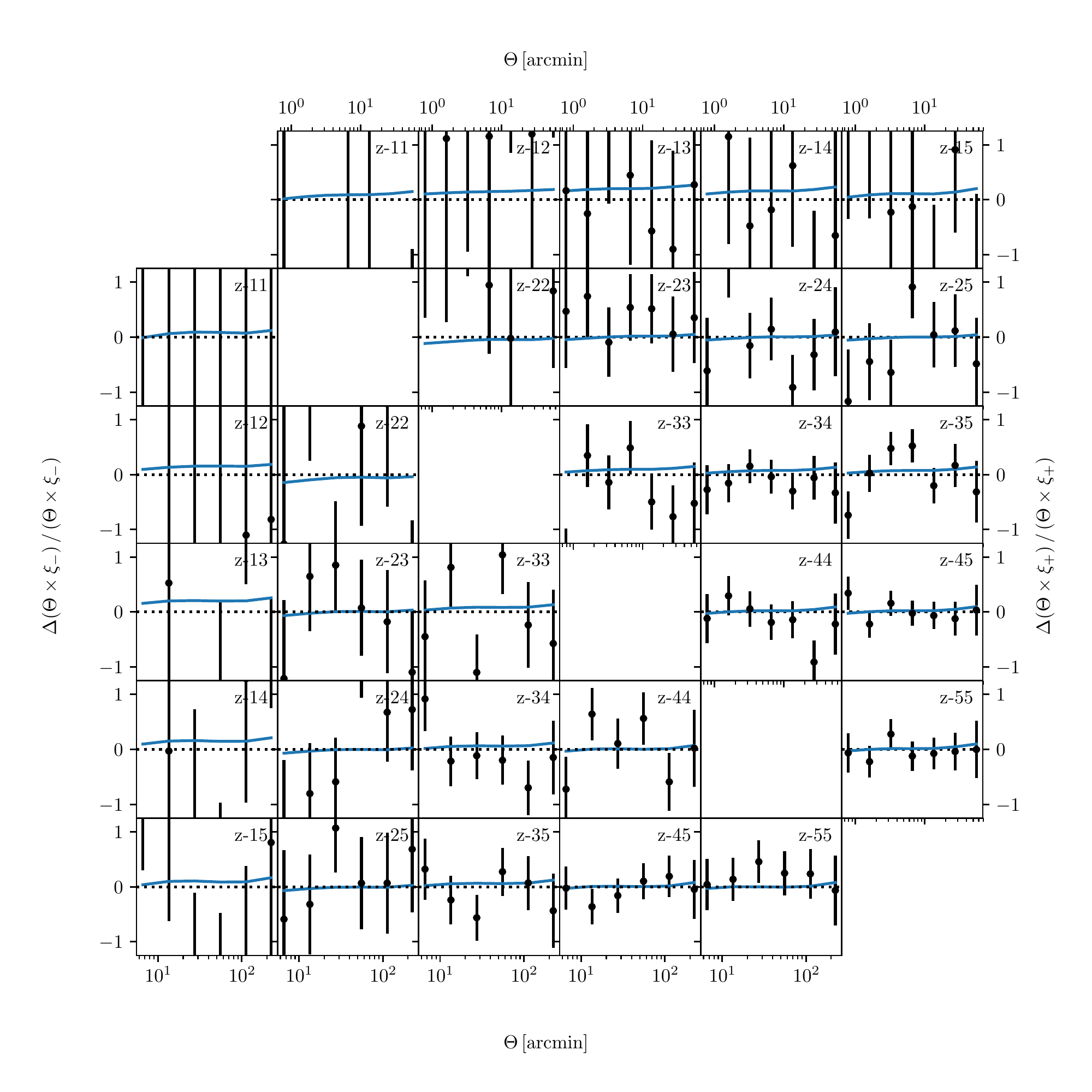}
\caption{Relative difference between the best-fit KV450 two-point shear correlation functions of the fiducial likelihood and the best fit after the iterative calibration of cosmological and nuisance parameters: $\xi_+$ (upper right) and $\xi_-$ (lower left). Black data points illustrate the relative difference between the observed two-point shear correlation functions and the best fit of the fiducial likelihood.}
\label{fig:correlation_function}
\end{figure*}

Furthermore, since the signal-to-noise ratio is lower in the low redshift bins compared to the high redshift bins, the relative contribution of intrinsic alignments is stronger in the low redshift bins. This can explain the larger shift in the median of the redshift distribution in the first two redshift bins that is observed in Fig. \ref{fig:medians_posterior}. This is illustrated in Fig. \ref{fig:correlation_function}, which shows the relative difference between the best-fit KV450 two-point shear correlation functions of the fiducial likelihood and the best fit after iterative calibration of cosmological and nuisance parameters. We find a relative difference of the best-fit curves of up to $20 \%$, which is largest in the first redshift bin. This is compatible with the observed shift in the median of the redshift distribution of the first redshift bin shown in Fig. \ref{fig:medians_posterior}. 

The black data points in Fig. \ref{fig:correlation_function} show the relative difference between the observed two-point correlation function and the fiducial best fit. These indicate that the signal-to-noise ratio in this bin is very low, so that the shift in the posterior redshift distribution does not have a significant impact on the overall best-fit likelihood value of the combined fit. In the second bin we find the smallest shift in the best-fit curve, although Fig. \ref{fig:medians_posterior} shows a significant shift in the median of the redshift distribution in this bin. This is an indication that the shift in the intrinsic alignment amplitude towards a lower value possibly mitigates the effect of the shifted redshift distribution, so that the effect on the likelihood value is minimal. From these observations, we conclude that the intrinsic alignment parameter $A_{\rm IA}$ does not solely measure the amplitude of intrinsic alignments, but instead picks up contributions from systematic shifts in the redshift distribution due to the degeneracy between the parameters. Variations between the constraints on the intrinsic alignment parameter were also reported by \cite{wright_som_kv450}, who found differences of up to $\left| \Delta A_{\rm IA} \right| \sim 1.0\sigma$ between analyses. However, since \cite{wright_som_kv450} used a different galaxy sample, the intrinsic alignment amplitude could be intrinsically different. Furthermore, the effect of the intrinsic alignment parameter mitigating systematic effects has been studied recently in other works, such as \cite{vanUitert18} and \cite{Efstathiou18}.

We conclude that our method provides constraints on cosmological parameters that are compatible with the fiducial KV450 analysis while taking all photometric redshift uncertainties into account. Our model provides a slightly better fit of the redshift distribution to the cosmic shear data since the large number of model parameters allows the model to reflect small variations in the redshift distribution. However, we suspect that variations in the redshift distribution can be correlated with variations in the intrinsic alignment amplitude, and therefore a tighter prior on the intrinsic alignment parameter through external constraints is required. Our approach can help reduce the degeneracy between intrinsic alignments and redshift distributions by providing a more accurate redshift distribution calibration.
\section{Summary and conclusions}
\label{sec:discussion}
In this paper we developed a method to model photometric redshift distributions of galaxy samples with strong correlations between tomographic bins using a modified Gaussian mixture model. We have shown that photometric redshift uncertainties arising from the calibration of the redshift distribution can be accurately propagated to the weak lensing likelihood via an analytic marginalisation over the model parameters.  This allowed us to use a fairly complex model of the redshift distribution without an increase in the number of sampling parameters in the weak lensing likelihood. Additionally, we developed an iterative method to fit cosmological and nuisance parameters in order to perform a self-calibration of the redshift distribution with cosmic shear data.

We applied these methods to the public KiDS+VIKING-450 (KV450) cosmic shear data. We fitted the modified Gaussian mixture model to the fiducial KV450 redshift distributions in five tomographic bins that were calibrated with deep spectroscopic surveys and implemented the marginalisation method in the public KV450 likelihood code. We performed the iterative fitting and found fast convergence to the best fit in the combined space of cosmological and nuisance parameters. Next, we sampled the weak lensing likelihood using the redshift distribution that was calibrated in the previous step and derived constraints on cosmological parameters.

We found that our model can fit complex redshift distributions thanks to the tunable number of model parameters. Since we marginalise analytically over nuisance parameters, the large number of redshift nuisance parameters does not increase the runtime of the posterior sampling. Our model provides a slightly better fit to the data compared to the fiducial KV450 likelihood since the fiducial likelihood only allows a shift in the mean of the redshift distribution of each bin and thus requires a pre-calibrated redshift distribution that closely resembles the true underlying distribution. Given the large uncertainties of photometric redshift calibration methods in general, a complex model that can reflect the uncertainties is advantageous. Therefore, with decreasing statistical uncertainties and increasing survey data, the method presented in this paper is particularly useful for upcoming surveys, where we expect higher order moments of the redshift distribution uncertainty to become increasingly important.

The marginalised posterior distributions of the remaining model parameters are in agreement with the fiducial KV450 analysis. However, we found slight shifts in the posterior constraints on the model parameters, which are strongest for the amplitude of intrinsic alignments, $A_{\rm IA}$. We suspect that these shifts are caused by degeneracies between the redshift distribution amplitudes and the intrinsic alignment amplitude, so that a shift in the redshift distribution can be compensated by a shift in the intrinsic alignment amplitude. This mitigation of systematic effects by the intrinsic alignment parameter is likely to be the reason for the relatively large shift in the median of the redshift distribution in the second redshift bin that we found after the iterative calibration of model parameters. Thus, to get unbiased constraints on the redshift distribution, we require a tighter prior on the intrinsic alignment parameter through external constraints. This will ensure that systematic effects are not absorbed by the intrinsic alignment parameter. This result is consistent with earlier works, such as \cite{wright_som_kv450}, \cite{hildebrandt18}, \cite{fortuna20}, \cite{vanUitert18}, and \cite{Efstathiou18}, which also found discrepant values of the intrinsic alignment amplitude and studied systematic effects on intrinsic alignments. Thus, this work further emphasises the necessity of an accurate modelling of intrinsic alignments.

While finalising this work, \cite{Hadzhiyska20} put forward a paper on the analytic marginalisation of redshift distribution uncertainties applied to galaxy clustering measurements from the HSC first data release. Their marginalisation method results in a modified data covariance matrix that downweights modes of the data vector that are sensitive to variations in the redshift distribution. This approach also allowed them to take the full shape of the redshift distribution into account. However, since this method directly modifies the data covariance matrix, it is unclear if it allows for a self-calibration of the redshift distribution with cosmic shear measurements. 

The method presented in this paper is not only applicable to cosmic shear analyses, but can also be adapted to other probes, such as galaxy-galaxy lensing and galaxy clustering. Therefore, it can especially be used in future joint `6x2pt' analyses, which combine all two-point correlation functions between overlapping imaging and spectroscopic surveys.
\begin{acknowledgements}
We thank the anonymous referee for their constructive comments, which helped to improve the manuscript.
We thank Catherine Heymans for helpful comments on this paper.
This work was partially enabled by funding from the UCL Cosmoparticle Initiative.
BJ and AHW acknowledge the kind hospitality of the Aspen Center for Physics, supported by National Science Foundation grant PHY-1607611, where this work was initiated.
AHW and HH are supported by an European Research Council Consolidator Grant (No. 770935).
HH is also supported by a Heisenberg grant of the Deutsche Forschungsgemeinschaft (Hi 1495/5-1). 
We acknowledge the use of data from the MICE simulations, publicly available at http://www.ice.cat/mice.
Based on data products from observations made with ESO Telescopes at the La Silla Paranal Observatory under programme IDs 177.A-3016, 177.A-3017, 177.A-3018, 179.A-2004, and 298.A-5015.
\end{acknowledgements}

\bibliographystyle{aa}
\bibliography{bibliography}

\begin{appendix} 
\section{Marginalisation over nuisance parameters}
\label{ap:derivatives}
In this appendix we provide the analytic expressions for the vector of derivatives of the log-likelihood with respect to the nuisance parameters $a_i^\alpha$ of the redshift distribution and the Hessian matrix of second derivatives with respect to the nuisance parameters. These quantities enter the calculation of the log-likelihood marginalised over the nuisance parameters described in Sect. \ref{sec:marginalisation}.
For the specific case of marginalising over the redshift distribution nuisance parameters, the vector $\boldsymbol{\cal L^\prime}$ has elements
\eqa{
\frac{\partial {\cal L}}{\partial a_m^\mu} &= \sum_{l,l',\alpha,\beta,\alpha',\beta'}  \left\{ \frac{\partial \Delta_l^{(\alpha \beta)}}{\partial a_m^\mu}\; Z_{(l,\alpha,\beta) \,  (l',\alpha',\beta')}\; \Delta_{l'}^{(\alpha' \beta')} \right. \\ \nn
& \hspace*{1.5cm} + \left. \Delta_l^{(\alpha \beta)}\; Z_{(l,\alpha,\beta) \,  (l',\alpha',\delta)}\; \frac{\partial \Delta_{l'}^{(\alpha' \beta')}}{\partial a_m^\mu} \right\}\;,
}
with
\eq{
\frac{\partial \Delta_l^{(\alpha \beta)}}{\partial a_m^\mu} = - A_m^\mu \sum_i x_\pm^{(im)}(\theta_l) \br{\delta_{\alpha \mu}\, A_i^\beta + \delta_{\beta \mu}\, A_i^\alpha}\;,
}
where $\delta_{\alpha \beta}$ denotes the Kronecker delta symbol. The indices $\alpha$ and $\beta$ run over all unique combinations of tomographic redshift bins. The two-point shear correlation function of two Gaussian comb components, $i$ and $j$, in $\theta$-bin $l$ is denoted by $x_\pm^{(ij)}(\theta_l),$ and the inverse data covariance is given by $\tens{Z}$. The difference between the observed and predicted signals, as defined in Eq. \eqref{eq:Delta}, is denoted by $\Delta_l^{\alpha\beta}$.
The elements of the Hessian matrix, ${\cal L^{\prime\prime}}$, read
\eqa{
\label{eq:hessian}
\frac{\partial^2 {\cal L}}{\partial a_m^\mu\, \partial a_n^\nu} &= \sum_{l,l',\alpha,\beta,\alpha',\beta'} 
\left\{ 
\frac{\partial \Delta_l^{(\alpha \beta)}}{\partial a_m^\mu}\; Z_{(l,\alpha,\beta) \,  (l',\alpha',\beta')}\; \frac{\partial \Delta_{l'}^{(\alpha' \beta')}}{\partial a_n^\nu}  \right. \\ \nn
&\hspace*{1.5cm}+ \frac{\partial \Delta_l^{(\alpha \beta)}}{\partial a_n^\nu}\; Z_{(l,\alpha,\beta) \,  (l',\alpha',\beta')}\; \frac{\partial \Delta_{l'}^{(\alpha' \beta')}}{\partial a_m^\mu} \\ \nn
&\hspace*{1.5cm}+ \frac{\partial^2 \Delta_l^{(\alpha \beta)}}{\partial a_m^\mu\, \partial a_n^\nu}\; Z_{(l,\alpha,\beta) \,  (l',\alpha',\beta')}\; \Delta_{l'}^{(\alpha' \beta')} \\ \nn
&\hspace*{1.5cm}\left. + \Delta_l^{(\alpha \beta)}\; Z_{(l,\alpha,\beta) \,  (l',\alpha',\beta')}\; \frac{\partial^2 \Delta_{l'}^{(\alpha' \beta')}}{\partial a_m^\mu\, \partial a_n^\nu}
\right\}\;,
}
with 
\eqa{
\frac{\partial^2 \Delta_l^{(\alpha \beta)}}{\partial a_m^\mu\, \partial a_n^\nu} &= - A_m^\mu A_n^\nu \, x_\pm^{(mn)}(\theta_l) \br{ \delta_{\alpha \mu} \delta_{\beta \nu} + \delta_{\beta \mu} \delta_{\alpha \nu} }\\ \nn
& ~~~~~ - \delta_{mn} \delta_{\mu \nu}\, A_m^\mu \sum_i x_\pm^{(im)}(\theta_l) \br{\delta_{\alpha \mu} A_i^\beta + \delta_{\beta \mu} A_i^\alpha}\;.
}
\section{Tests of the redshift distribution calibration}
\label{ap:calibration}
In this appendix we test to what extent the number of Gaussian components of the comb model affects the fit to the pre-calibrated redshift histograms. Additionally, we test the stability of the fit results when changing the number of bins of input data histograms and test the calibration method on simulations.
\subsection{Width of input data histograms}
\label{ap:calibration_input}
The redshift distributions of \cite{hildebrandt18}, calibrated with the fiducial DIR method, consist of histograms with bin width $\Delta z = 0.05$ for each tomographic bin and a covariance matrix that links all five tomographic bins. To test the sensitivity of the fit with respect to the input data, we performed fits using a second set of histograms with a smaller bin width of $\Delta z = 0.025$ that were calibrated with the same method. 

\begin{figure*}
\centering
\includegraphics[width=\linewidth]{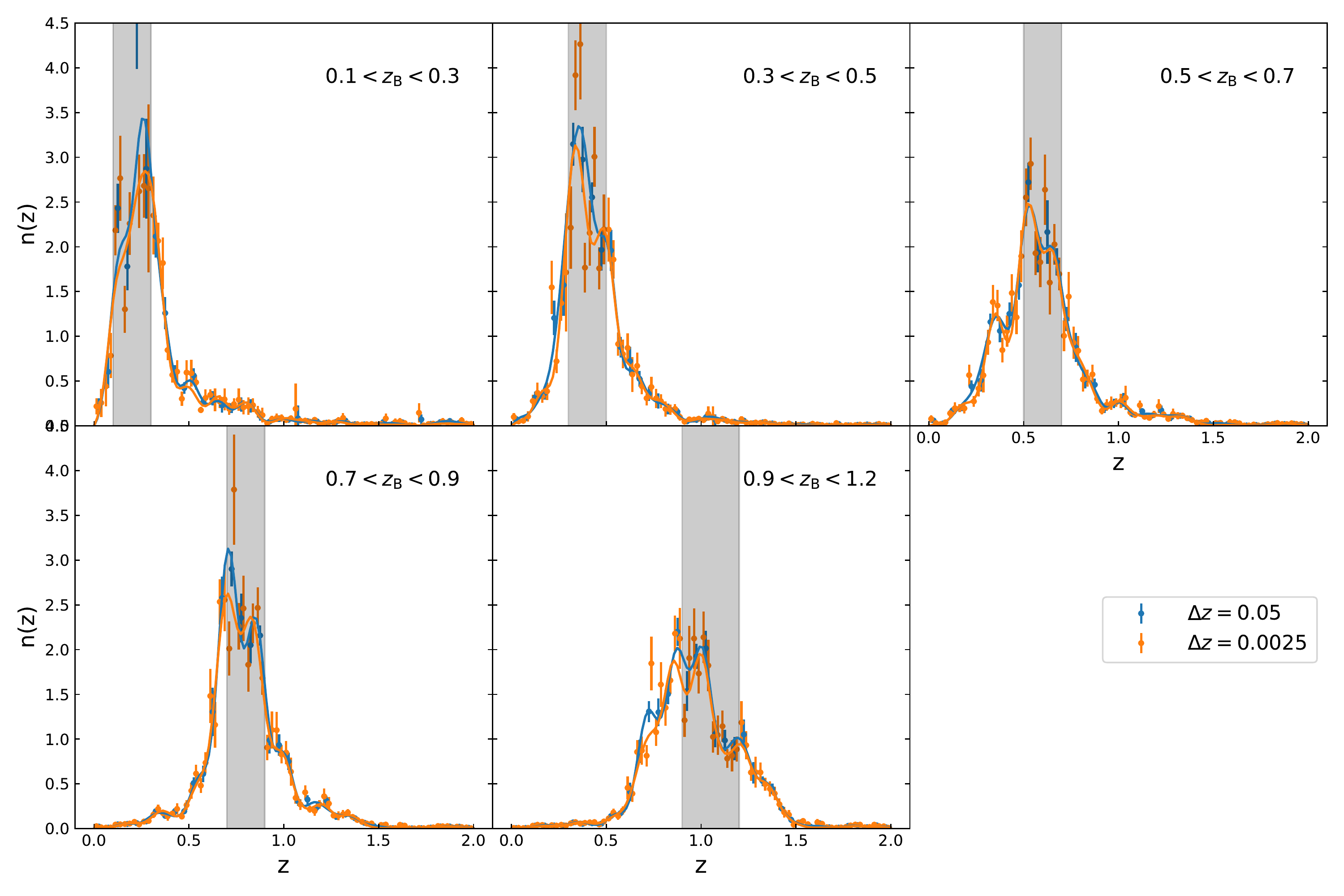}
\caption{Comparison of the Gaussian comb with 30 components fitted to two different pre-calibrated histograms. The blue and orange points show histograms with bin widths of $\Delta z = 0.05$ and $\Delta z = 0.025$, respectively. The error bars correspond to the diagonal elements of the covariance matrix. The lines represent the Gaussian comb with 30 components fitted to the data histograms. We note that when fitting the redshift distribution, the full covariance matrix of the data histogram is taken into account.}
\label{fig:120vs240}
\end{figure*}

Figure \ref{fig:120vs240} shows a comparison of two fits with 30 Gaussian components; the blue lines represent a fit to the histograms with bin width $\Delta z = 0.05$, and the orange lines represent a fit to the histograms with bin width $\Delta z = 0.025$. We note that the error bars correspond to the diagonal elements of the covariance matrix of the data histograms. The fit of redshift distributions, however, is performed using the full covariance matrix.

By visually inspecting the fitted redshift distributions, we observe some deviations between the two curves, which are, however, already present in the input data. Although the two histograms are supposed to represent the same source redshift distribution, they show some fluctuations (especially in the peaks of the distributions), which have an impact on the fitted curves. More importantly, however, we find goodness of fit values of $\chi^2 = 4500$ and $\chi^2 = 22750$ for 50 and 250 degrees of freedom, respectively. This indicates a bad fit of the model to the data regardless of which data histogram is used. To find the cause of the bad fit, we repeated the fit using only the diagonal elements of the covariance matrix, which reduces the goodness of fit values to $\chi^2 = 470$ and $\chi^2 = 1400$ for 50 and 250 degrees of freedom, respectively. This test shows that the bad fit is to a great extent caused by the off-diagonal elements of the covariance matrix. However, excluding the off-diagonal elements does not lead to an acceptable goodness of fit value. We suspect that the uncertainties on the pre-calibrated redshift distributions are underestimated, which would explain the discrepancies between the blue and orange data points shown in Fig. \ref{fig:120vs240}.

Empirically, we find that rescaling the square root of the covariance $C_{ij}$ between histogram bins by an additive and multiplicative factor via
\eq{
C^\prime_{ij} = \left(2\sqrt{C_{ij}}+0.01\delta_{ij}\right)^2
}
leads to $\chi^2 = 80$ for 50 degrees of freedom. With this rescaling, the widths of the posterior distributions of the median redshift for both the redshift histograms and the Gaussian comb, shown in Fig. \ref{fig:comb}, inflate by approximately the same factor. Therefore, we assume that a potential underestimation of the error bars impacts the fiducial analysis and the analysis presented in this paper in the same way. We stress that the quality of the fit does not have a significant impact on the main analysis of this paper, as shown in Appendix \ref{sec:MICE}, and leave further investigation of an improved uncertainty quantification for the pre-calibrated redshift distributions for future work.
\subsection{Redshift distribution calibration with simulations}
\label{sec:MICE}
\begin{figure*}
\centering
\includegraphics[width=\linewidth]{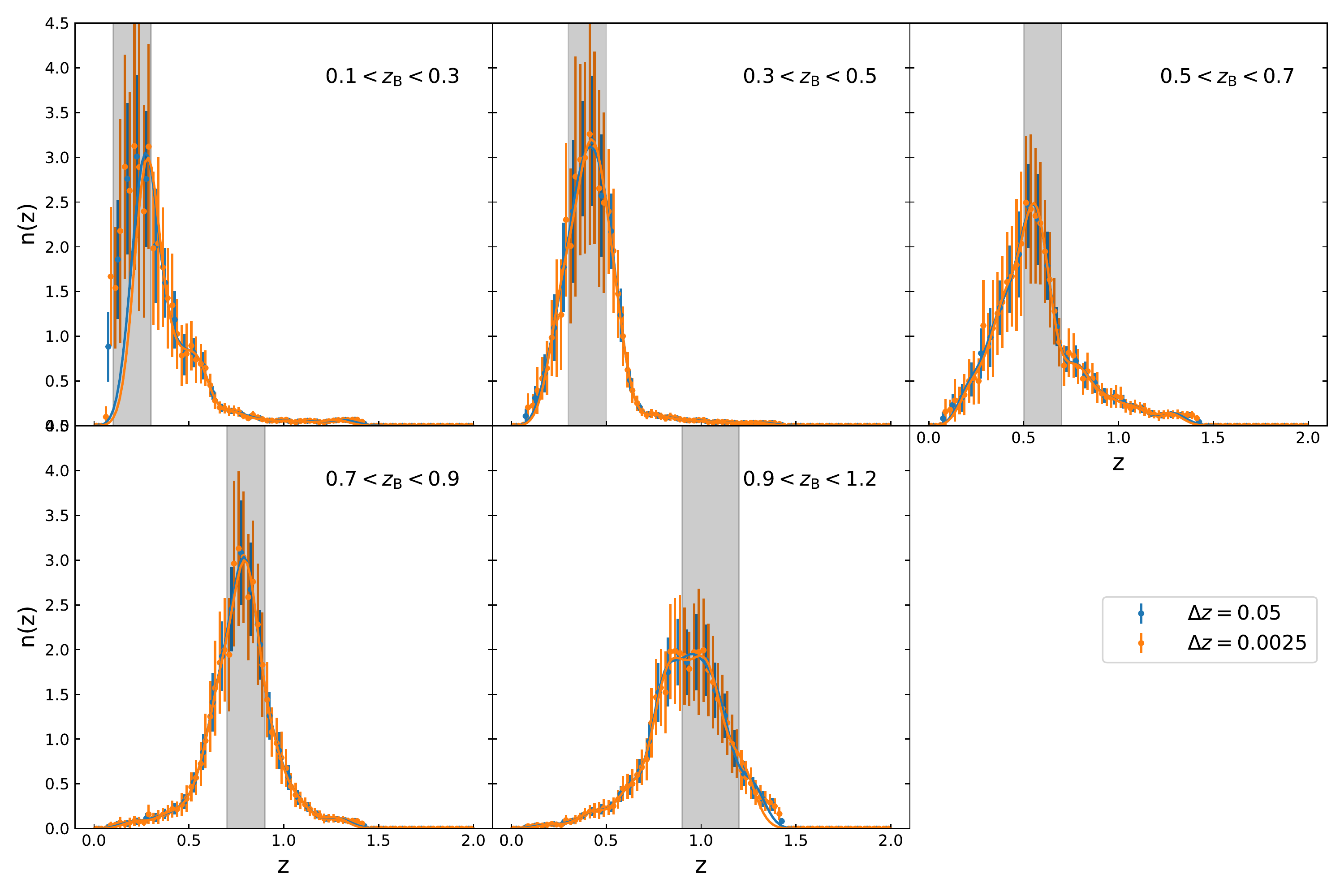}
\caption{Comparison of the Gaussian comb with 30 components fitted to pre-calibrated histograms from the MICE simulation. The blue and orange points show histograms with bin widths of $\Delta z = 0.05$ and $\Delta z = 0.025$, respectively. The error bars correspond to the diagonal elements of the covariance matrix. The lines represent the Gaussian comb with 30 components fitted to the data histograms. We note that when fitting the redshift distribution, the full covariance matrix of the data histogram is taken into account.}
\label{fig:MICE}
\end{figure*}

Since in Appendix \ref{ap:calibration_input} we found that the Gaussian comb model provides a bad fit to the actual data, which is likely caused by an underestimation of the error bars, we tested our calibration method with simulations. We used redshift distributions that are calibrated with the fiducial DIR method on simulated mock catalogues \citep{vdBusch20} based on the MICE simulation \citep{Mice1, Mice2, Mice3, carretero, hoffmann}. Analogous to Appendix \ref{ap:calibration_input}, we compared two types of data histograms with bin widths of $\Delta z = 0.05$ and $\Delta z = 0.025$. Figure \ref{fig:MICE} shows a comparison of the two fits with 30 Gaussian components, where blue lines represent a fit to the histograms with bin width $\Delta z = 0.05$ and orange lines represent a fit to the histograms with bin width $\Delta z = 0.025$. 

We find that with goodness of fit values of $\chi^2 = 75$ and $\chi^2 = 320$ for 50 and 250 degrees of freedom, respectively, our Gaussian comb model fits the data reasonably well. Moreover, both the data histograms and the corresponding fitted redshift distributions are in excellent agreement. We conclude that the Gaussian comb model is capable of accurately describing the redshift distribution. The worse goodness of fit when fitting real data is likely due to the presence of noise and an underestimation of the uncertainties. 
\subsection{Number of Gaussian components}
\begin{figure*}
\centering
\includegraphics[width=\linewidth]{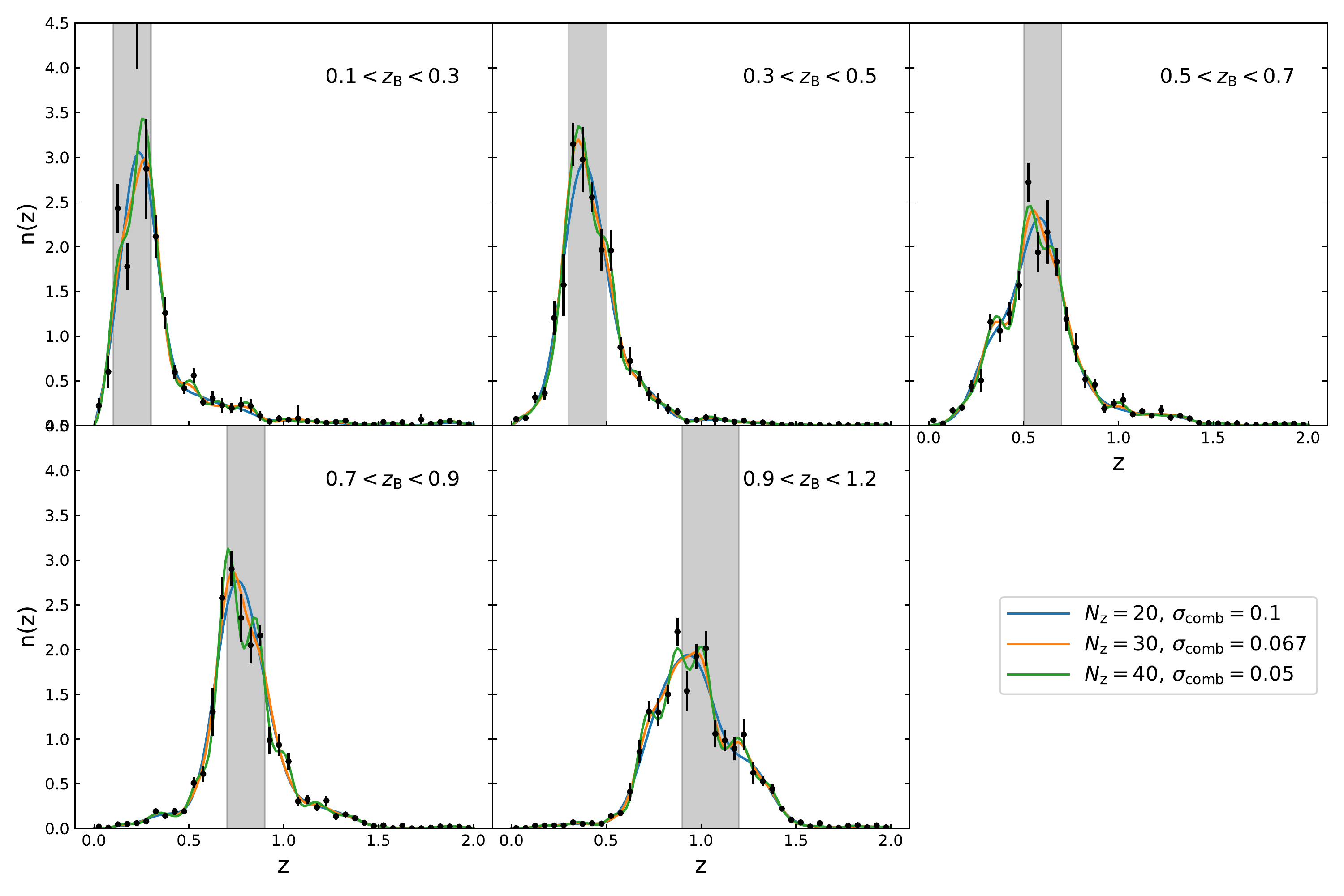}
\caption{Comparison of a Gaussian comb with 20, 30, and 40 components fitted to a pre-calibrated histogram with bin width $\Delta z = 0.05$. The width $\sigma_{\rm comb}$ of the Gaussians is equal to the separation between each component. Data points are shown in black, with error bars corresponding to the diagonal elements of the covariance matrix. Blue, orange, and green lines represent the Gaussian combs with 20, 30, and 40 components, respectively. We note that when fitting the redshift distribution, the full covariance matrix of the data histogram is taken into account.}
\label{fig:n_comp}
\end{figure*}

In Fig. \ref{fig:n_comp} we show a comparison of fits of a Gaussian comb with 20, 30, and 40 components to the pre-calibrated redshift histograms with bin width $\Delta z = 0.05$.  The width $\sigma_{\rm comb}$ of the Gaussians is equal to the separation between each component. We find that variations in the number of comb components have a marginal impact on the redshift distribution, with changes of the median of order $\Delta z^{\rm med} = 0.001$.
\section{Comparison between the fiducial KV450 likelihood and the modified likelihood with Gaussian comb}
\label{ap:kv450_likelihood}
In order to test if the fitted redshift distribution is capable of reproducing the results of the fiducial KV450 analysis, we sampled the likelihood using the Gaussian comb model as the parameterisation of the redshift distribution, but without marginalisation over the uncertainties on the nuisance parameters. To be able to compare the two likelihoods, we fixed the nuisance parameters $\delta z_i$ of the fiducial KV450 likelihood. The results of these fits are presented Table \ref{tab:consistency}, which shows the mean posterior values of cosmological and nuisance parameters. We find that constraints from both setups are fully consistent, and therefore we conclude that our Gaussian comb model can be used as an alternative to the fiducial redshift distributions.

\begin{table}
\caption{Comparison between the fiducial KV450 likelihood and the modified likelihood with redshift distribution parameterised by the Gaussian comb model.}
\centering
\begin{tabular}{lll}
\hline\hline
Parameter & fiducial KV450 & KV450 with Gaussian comb\\
\hline
$\omega_{\rm cdm }  $ & $0.112^{+0.029}_{-0.060}   $& $0.112^{+0.046}_{-0.060}   $\\

$\ln10^{10}A_{\rm s}$ & $3.30\pm 0.92              $& $3.34\pm 0.92              $\\

$\omega_{\rm b}    $ & $0.0223\pm 0.0021          $& $0.0222^{+0.0018}_{-0.0025}$\\

$n_{\rm s}         $ & $1.03^{+0.15}_{-0.13}      $& $1.01\pm 0.13              $\\

$h              $ & $0.749^{+0.067}_{-0.028}   $& $0.746^{+0.062}_{-0.033}   $\\
\hline
$A_{\rm IA}        $ & $0.89^{+0.64}_{-0.58}      $& $0.87^{+0.64}_{-0.58}              $\\

$c_{\rm min}       $ & $2.50^{+0.22}_{-0.45}      $& $2.49^{+0.23}_{-0.40}     $\\

$\delta c             $ & $0.00000\pm 0.00019        $& $0.00000\pm 0.00019        $\\

$A_c             $ & $1.03\pm 0.12              $& $1.02\pm 0.12              $\\

$\Omega_{\rm m}    $ & $0.242^{+0.052}_{-0.11}    $& $0.242^{+0.055}_{-0.11}    $\\

$\sigma_8        $ & $0.86^{+0.18}_{-0.20}      $& $0.87\pm 0.17      $\\
$S_8$           & $0.746_{-0.028}^{+0.029}$&$0.748_{-0.03}^{+0.029}$\\
\hline
\end{tabular}
\tablefoot{Reported are the mean posterior values and the 68\% confidence intervals. The first five lines are cosmological parameters, and the remaining lines represent nuisance parameters.}
\label{tab:consistency}
\end{table}

\section{Posteriors of cosmological parameter constraints}
\label{ap:posteriors}
\begin{figure*}
\centering
\includegraphics[width=\linewidth]{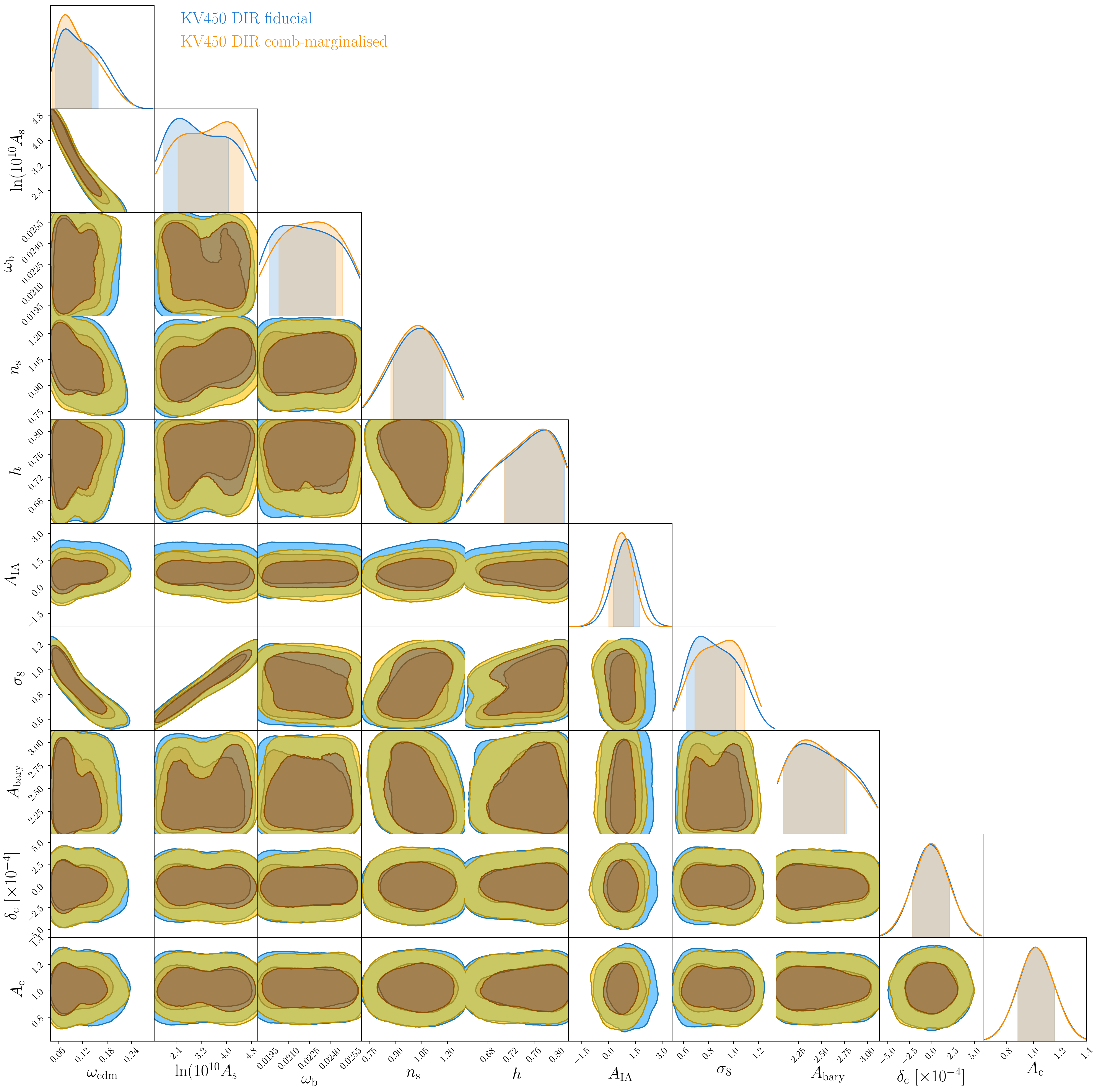}
\caption{Marginalised posteriors for all parameters of the KV450 likelihood. Blue contours present the results from the KV450 likelihood with a Gaussian comb and analytical marginalisation over nuisance parameters, while the orange contours refer to the fiducial KV450 constraints.}
\label{fig:posteriors}
\end{figure*}

In Fig. \ref{fig:posteriors} we show marginalised posteriors of cosmological and nuisance parameters. The KV450 likelihood with a Gaussian comb and analytical marginalisation over nuisance parameters is compared to the fiducial KV450 likelihood.
\end{appendix}
\end{document}